\documentclass[fleqn,usenatbib]{mnras}

\usepackage{newtxtext,newtxmath}

\usepackage[T1]{fontenc}
\usepackage{ae,aecompl}
\usepackage{siunitx}
\usepackage[para]{threeparttable}
\usepackage{longtable}
\usepackage{tabularray}
\usepackage{subfigure}

\usepackage{graphicx}	
\usepackage{amsmath}	
\usepackage{wrapfig,lipsum}
\usepackage{hyperref}
\usepackage{upgreek}

\newcommand{\FRB}{FRB\,20240619D}
\newcommand{\pccm}{\,pc\,cm$^{-3}$}

\title[FRB search]{MeerKAT discovery of a hyperactive repeating fast radio burst source}

\author[J. Tian et al.]{J. Tian$^{1}$\thanks{E-mail: jun.tian@manchester.ac.uk}, I. Pastor-Marazuela$^1$\thanks{E-mail: ines.pastor.marazuela@gmail.com}, K. M. Rajwade$^2$, B. W. Stappers$^1$, K. Shaji$^{5,6}$,
\newauthor K. Y. Hanmer$^{9,10}$, M. Caleb$^{5,6}$, M. C. Bezuidenhout$^{3,4}$, F. Jankowski$^7$, R. Breton$^{1}$, 
\newauthor E. D. Barr$^{8}$, M. Kramer$^8$, P. J. Groot$^{9,11,12}$, S. Bloemen$^{12}$, P. Vreeswijk$^{12}$, D. Pieterse$^{12}$,
\newauthor P. A. Woudt$^9$, R. P. Fender$^{2,9}$, R. A. D. Wijnands$^{13}$, D. A. H. Buckley$^{9,11}$
\\
$^1$Jodrell Bank Centre for Astrophysics, Department of Physics and Astronomy, The University of Manchester, Manchester M13 9PL, UK\\
$^2$Astrophysics, The University of Oxford, Denys Wilkinson Building, Keble Road,
Oxford OX1 3RH, UK\\
$^3$Centre for Space Research, North-West University, Potchefstroom 2531, South Africa\\
$^{4}$Department of Mathematical Sciences, University of South Africa, Cnr Christiaan de Wet Rd and Pioneer Avenue,\\
Florida Park, 1709, Roodepoort, South Africa\\
$^5$Sydney Institute for Astronomy, School of Physics, The University of Sydney, NSW 2006, Australia\\
$^6$ARC Centre of Excellence for Gravitational Wave Discovery (OzGrav), Hawthorn, 3122, Victoria, Australia\\
$^7$LPC2E, OSUC, Univ Orleans, CNRS, CNES, Observatoire de Paris, F-45071 Orleans, France\\
$^8$Max-Planck-Institut fur Radioastronomie, 53121 Bonn, Germany\\
$^9$Department of Astronomy, University of Cape Town, Private Bag X3, Rondebosch 7701, South Africa\\
$^{10}$High Energy Physics, Cosmology \& Astrophysics Theory (HEPCAT) Group, Department of Mathematics \& Applied Mathematics,\\
University of Cape Town, Cape Town, 7700, South Africa\\
$^{11}$South African Astronomical Observatory, P.O. Box 9, Observatory, 7935, South Africa\\
$^{12}$Department of Astrophysics/IMAPP, Radboud University, P.O. Box 9010, 6500 GL, Nijmegen, The Netherlands\\
$^{13}$Anton Pannekoek Institute for Astronomy, University of Amsterdam, P.O. Box 94249, 1090 GE Amsterdam, The Netherlands
}%

\date{Accepted XXX. Received YYY; in original form ZZZ}

\pubyear{2024}

\begin{document}
\label{firstpage}
\pagerange{\pageref{firstpage}--\pageref{lastpage}}
\maketitle

\begin{abstract}
We present the discovery and localisation of a repeating fast radio burst (FRB) source from the MeerTRAP project, a commensal fast radio transient search programme using the MeerKAT telescope. \FRB\ was first discovered on 2024 June 19 with three bursts being detected within two minutes in the MeerKAT L-band (856--1712\,MHz). We conducted follow-up observations of \FRB\ with MeerKAT using the Ultra-High Frequency (UHF; $544\text{--}1088$\,MHz), L-band and S-band (1968--2843\,MHz) receivers one week after its discovery, and recorded a total of 249 bursts.
The MeerKAT-detected bursts exhibit band-limited emission with an average fractional bandwidth of 0.31, 0.34 and 0.48 in the UHF, L-band and S-band, respectively. We find our observations are complete down to a fluence limit of $\sim1$\,Jy\,ms, above which the cumulative burst rate follows a power law $R (>F)\propto (F/1\,\text{Jy}\,\text{ms})^\gamma$ with $\gamma=-1.6\pm0.1$ and $-1.7\pm0.1$ in the UHF and L-band, respectively. 
The near-simultaneous L-band, UHF and S-band observations reveal a frequency dependent burst rate with $3\times$ more bursts being detected in the L-band than in the UHF and S-band, suggesting a spectral turnover in the burst energy distribution of \FRB. Our polarimetric analysis demonstrates that most of the bursts have $\sim100\%$ linear polarisation fractions and $\sim10\%\text{--}20\%$ circular polarisation fractions. We find no optical counterpart of \FRB\ in the MeerLICHT optical observations simultaneous to the radio observations and set a fluence upper limit in MeerLICHT's $q$-band of 0.76\,Jy\,ms and an optical-to-radio fluence ratio limit of 0.034 for a 15\,s exposure.
\end{abstract}




\section{INTRODUCTION}

Fast radio bursts (FRBs) are short-duration ($\mu\text{s}\text{--}\text{ms}$; e.g. \citealt{Cho20, Nimmo21}), highly energetic ($\sim10^{35}\text{--}10^{42}$\,erg; \citealt{Ryder23}) bursts of coherent radio emission arising from unknown extra-galactic sources \citep{Cordes19, Petroff22}. Thousands of FRB sources have been discovered so far, mostly by the Canadian Hydrogen Intensity Mapping Experiment (CHIME; \citealt{CHIME22}) Fast Radio Burst project (CHIME/FRB; \citealt{CHIME18}), with the majority observed as one-off events \citep{Chime23}. A small fraction of FRBs ($\sim3\%$) are observed to repeat and show statistically different burst properties from the apparent non-repeaters, such as wider pulse widths and narrower emission bandwidths \citep{Pleunis21b}. However, it remains unclear whether repeating and non-repeating FRBs share the same origin \citep{Caleb19, Chime23}.

Repeating FRBs show a wide range of repetition rates. Some sources have only two bursts detected with burst rates comparable to the rate upper limits of some apparent non-repeaters \citep{Chime23, Kirsten24}. 
On the other end, a few active repeating sources are found to emit tens to hundreds of bursts per hour, such as FRB~20121102A \citep{Li21, Jahns23}, FRB~20201124A \citep{Lanman22} and FRB~20200120E \citep{Nimmo23}. Their bursting activity is often clustered in time. 
FRB~20121102A and FRB~20180916B show periodic clusters of activity with bursts being detectable only within an active window in each cycle \citep{Rajwade20c, CHIME20b, Cruces21}, which might reflect an orbital \citep{Ioka20}, rotational \citep{Beniamini20} or precession \citep{Levin20} period. Other active repeaters show burst storms where the source suddenly switches on from quiescence and enters a hyperactive state with a burst rate of up to hundreds per hour that lasts for weeks to months, e.g., FRB~20201124A \citep{Lanman22, Xu22, Bilous24}, FRB~20220912A \citep{Feng23, Hewitt23, Zhang23b} and FRB~20240114A \citep{Panda24, Kumar24, Tian24b}. The triggering mechanism of this high activity is still unknown, and has been suggested to be starquakes or magnetar flares (e.g., \citealt{Wang18, Totani23}). Studying more such hyperactive repeaters, especially the statistical properties of their burst occurrence times and energies, can further test these models.

Repeating FRBs often show complex burst morphologies (e.g., \citealt{CHIME19b, Fonseca20, Chime23}). One prominent feature is the downward drifting of sub-bursts in frequency with time \citep{Hessels19, CHIME19c, Caleb20, Pleunis21b}. This may be simply explained with radius-to-frequency mapping in the context of FRB emission arising from charged particles within the magnetosphere of a neutron star \citep{Wang19}. Microstructures have also been observed in some active repeaters. FRB~20180916B shows sub-bursts as narrow as $30\,\mu$s at 800\,MHz \citep{Sand22} and $3\,\mu$s at 1.7\,GHz \citep{Nimmo21}. Even narrower sub-bursts - on the order of 60\,ns - were seen from FRB~20200120E at 1.4\,GHz \citep{Nimmo22} and 2.3\,GHz \citep{Majid21}. Such microstructures resemble nano-shots from the Crab pulsar \citep{Hankins03} and might suggest that the FRB emission is magnetically powered \citep{Philippov19, Lyubarsky20, Lyutikov21}. High-time-resolution studies of more active repeaters (e.g., \citealt{Hewitt23, Snelders23, Nimmo23}) could reveal more such microstructures and further constrain the FRB emission mechanism.

A diversity of polarimetric properties have been observed among FRBs. Most FRBs possess high linear polarisation fractions up to $\sim100\%$ (e.g., \citealt{Gajjar18, Cordes19, CHIME19b, Fonseca20}; \citealt{Zhang23b}; \citealt{Pandhi24}), while some exhibit significant circular polarisation (e.g., \citealt{Jiang22, Kumar22, Xu22}). There are diverse patterns of polarisation position angle (PPA) variations across each burst, including different PPA swings in FRB~20180301A \citep{Luo20}, S-shaped evolution in FRB~20221022A \citep{Mckinven24} and orthogonal jumps in FRB~20201124A \citep{Niu24}, while most other FRBs show a flat PPA (e.g., \citealt{Michilli18, Day20, Kumar21c}). These observations suggest magnetospheric origins for at least some FRBs. Frequency-dependent linear polarisation has been observed in some repeating FRBs, considered to be caused by multipath propagation in the inhomogeneous magneto-ionic environment of the FRB source \citep{Feng22, Mckinven23}. A sign change in the rotation measure (RM) has also been reported for two repeating FRBs, 20190520B and 20180301A, suggesting magnetic field reversal in the FRB environment \citep{Anna-Thomas23, Kumar23}. These polarimetric features have started to emerge and shed light on the complexity of the magnetised environments associated with repeating FRBs.

On 2024 June 19, \FRB\ was discovered by MeerKAT during the MeerTime Thousand Pulsar Array (TPA) observation (proposal id: SCI-20180516-MB-02; PI: Simon Johnston), with three bursts observed within 2\,min in the L-band between 856--1712\,MHz \citep{Tian24c}. We measured a signal-to-noise (S/N) maximising dispersion measure (DM) of 486.8\pccm. The bursts triggered channelised voltage data from the individual MeerKAT antennas, which we used to initially localise the FRB in the image domain 
using our transient buffer imaging pipeline \citep{Rajwade20, Rajwade24}. 
No bursts were identified in two archival data sets from 2024.
This suggests that \FRB\ could have entered a sudden period of high activity. Therefore, we were motivated to propose for Director's Discretionary Time to follow up with MeerKAT (proposal id: DDT-20240625-JT-01), which was allocated 2\,hr of observation on 2024 June 26. Subsequent detections of many more bursts by MeerKAT, the Westerbork RT-1 25\,m telescope \citep{Ould24} and the upgraded Giant Metrewave Radio Telescope (uGMRT; \citealt{Kumar24b}) confirmed the hyperactivity of \FRB. In order to study the evolution of the spectral, temporal and polarimetric properties of \FRB\ over a wide and continuous frequency band, we carried out a further 2\,hr follow-up using the MeerKAT sub-array mode where half of the antennas observed at Ultra-High Frequency (UHF; 544--1088\,MHz) and the other half at L-band simultaneously. This also allowed us to study the band-limited nature of individual bursts.

In this work, we report the discovery of this new hyperactive repeater, \FRB, with MeerKAT, and the detection of 249 bursts from further follow-up. In Section~\ref{sec:observation}, we describe the observational configuration of MeerKAT and the transient detection pipeline. We also describe MeerLICHT contemporaneous optical observations aimed to search for optical counterparts of \FRB. Our results are then presented in Section~\ref{sec:results}. We discuss the implications of our results for FRB sources and emission models in Section~\ref{sec:discussion}, followed by conclusions in Section~\ref{sec:conclusions}.

\section{Observations and data}\label{sec:observation}

We discovered \FRB\ at 00:41 UTC on 2024 June 19 as part of the MeerTRAP (More TRansients And Pulsars) project, a commensal fast radio transient-detection programme that piggybacks other large survey projects using MeerKAT in South Africa \citep{Sanidas18, Bezuidenhout22, Rajwade22, Caleb23, Jankowski23, Driessen24}. At that time, MeerKAT was operating in a sub-array mode, with 32 of the 64 13.5-m dishes included in the sub-array which detected the FRB. All three bursts were detected in the L-band within 2\,min and triggered channelised voltage data. 
We used these data to localise the source (see Section~\ref{sec:imaging}).

Given the limited exposure of MeerKAT at the source position ($\sim2$\,min on 2024 June 19), \FRB\ appeared to be highly active. Therefore, we carried out a 2\,hr follow-up observation with MeerKAT, 1\,hr at L-band and 1\,hr at S-band ($1968\text{--}2843$\,MHz), starting on 2024 June 26 at 00:09 UTC as part of DDT-20240625-JT-01. We used the inner 40 dishes of the $\sim$1-km core of the array. The primary beam full width at half-maximum (FWHM) at the L-band and S-band are $\sim1.3\,\text{deg}^2$ and $\sim0.4\,\text{deg}^2$, respectively \citep{Mauch20}, sufficient for covering the positional error region of \FRB. Therefore, we recorded imaging data for the FRB source localisation as well as beamformed data to search for repeat bursts.

We deployed the Transient User Supplied Equipment (TUSE), a real-time transient detection backend instrument developed by MeerTRAP, to trigger voltage buffer dumps while searching for bursts in real-time. We observed in the coherent beamforming mode, where voltages from the inner 40 dishes were coherently combined and phased using the Filterbank and Beamforming User Supplied Equipment (FBFUSE), a multi-beam beamformer 
\citep{Barr18, Chen21}. We used the typical MeerTRAP observing configuration with 768 tied-array coherent beams (CBs) overlapping at 75\% maximum and tiling out from the initial localisation of \FRB. Data from these CBs were arranged and ingested by the TUSE real-time single pulse search pipeline. 
Subsequent detections triggered channelised, high time resolution transient buffer data to be saved for offline correlation and imaging (for the MeerTRAP voltage buffer dump system see \citealt{Rajwade24}). This observing strategy produced two datasets: detected time-frequency data in the filterbank format from all CB detections and voltage data for all triggers. Each filterbank file contains a dispersed pulse and additional padding of 0.5\,s at the start and end of the file. It has 1024 frequency channels and a frequency and time resolution of 0.84\,MHz/306.24\,$\mu$s and 0.85\,MHz/449.39\,$\mu$s at L-band and S-band, respectively. The voltage data are 300\,ms long for each trigger and incoherently dedispersed at the detection DM before being saved to disc. They are Nyquist sampled across 4096 channels from 62 out of 64 dishes available at that time.

We performed additional follow-up of the \FRB\ source using the MeerKAT sub-array mode on 2024 June 28 from 00:43 UTC to 02:43 UTC. The MeerKAT dishes were equally split into two sub-arrays (32 dishes in each) with similar synthesised beams, allowing us to observe simultaneously at UHF and L-band with similar sensitivities at the FRB position.
As the TUSE backend only works for one sub-array, we used it with the sub-array observing at UHF, and for the other sub-array at L-band we used the Pulsar Timing User-Supplied Equipment (PTUSE; \citealt{Bailes20}) to coherently beamform at the position of \FRB. We chose to run PTUSE in the 4k channel search mode with a frequency and time resolution of 0.21\,MHz/38.28\,$\mu$s and full Stokes information. The resulting PTUSE data were saved to disc and then searched offline for repeat bursts.

\subsection{Burst detection}\label{sec:detection}

\begin{figure*}
\centering
\includegraphics[width=.85\textwidth]{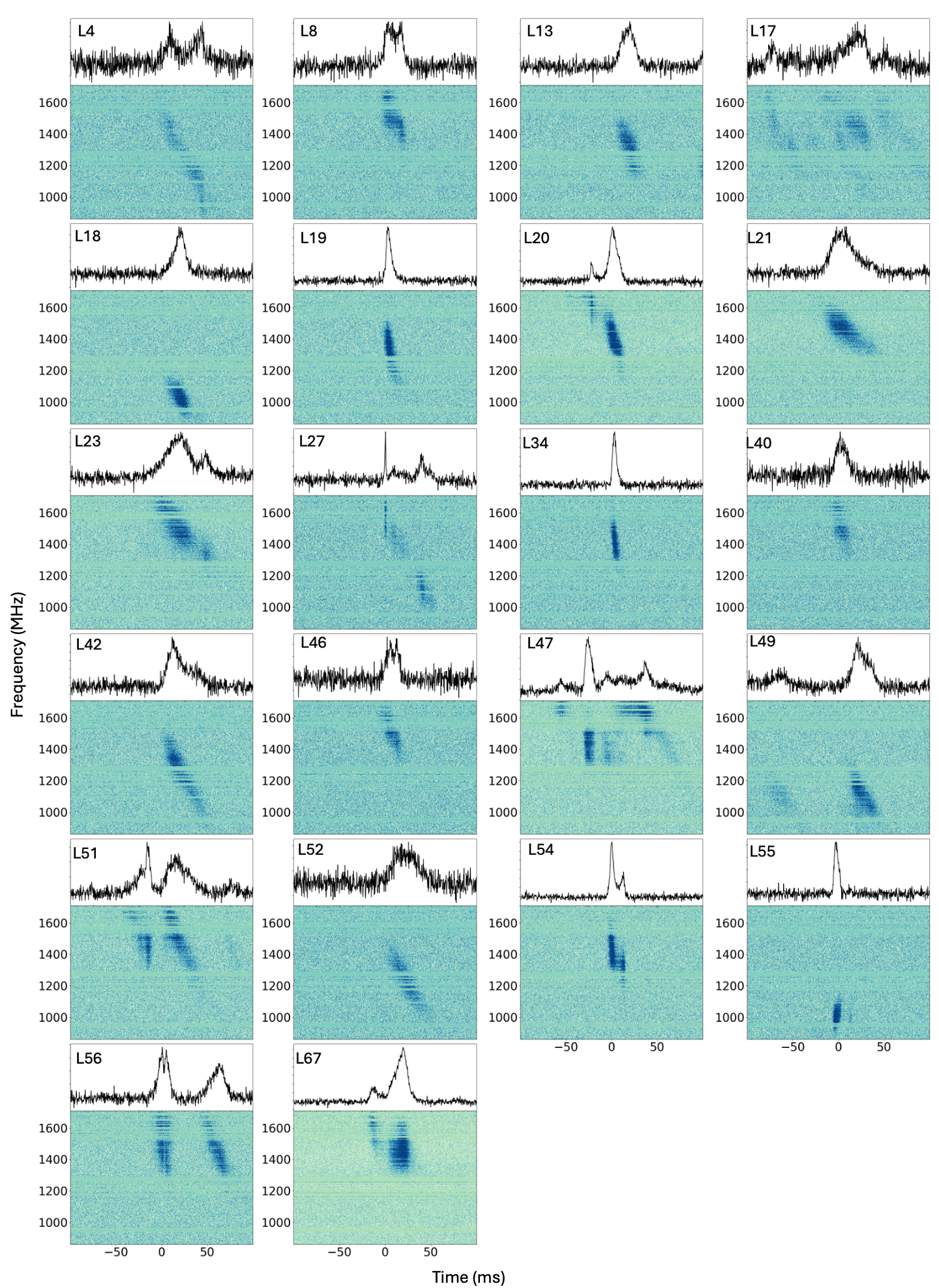}
\caption{A subset of the bursts detected from \FRB\ in chronological order. This sample consists of very bright bursts or those showing multiple sub-components and/or complex time-frequency structure. Each panel shows the dynamic spectrum of a burst from Table~\ref{tab:bursts} from the filterbank data dedispersed to $\text{DM}=464.85$\pccm, the structure optimising DM determined for \FRB\ in Section~\ref{sec:DM}, with the top sub-panel showing the frequency-averaged pulse profile in arbitrary units. The dynamic spectra have been binned $4\times$ in frequency. A label is given to each burst in the top-left corner with "U", "L" and "S" indicating detection in the UHF, L-band and S-band, respectively. Burst L49 follows L48 closely, and part of the L48 emission is visible in the panel of L49. The horizontal lines that show the same colour are either missing channels or flagged due to RFI.}
\label{fig:bursts}
\end{figure*}

We searched for bursts using two methods: a real-time search using the state-of-the-art GPU-based single pulse search pipeline {\sc astroaccelerate}\footnote{\url{https://github.com/AstroAccelerateOrg/astro-accelerate}} \citep{Armour12, Adamek20} on the TUSE data, including the full-array L-band and S-band observations and the sub-array UHF observation, and an offline search using the GPU-based single-pulse search software {\sc heimdall}\footnote{\url{https://sourceforge.net/projects/heimdall-astro/}} \citep{Barsdell12} on the PTUSE data, i.e. the sub-array L-band observation. In the real-time search, all candidates with $\text{S/N}>8$ were saved to disc in the filterbank format. More details on removing radio frequency interference (RFI) and sifting the candidates can be found in \citet{Caleb22}, \citet{Rajwade22} and \citet{Jankowski23}. After visually inspecting the pulse profiles and dynamic spectra of the candidates, we obtained a total of 141 bursts, 69 from the 1\,hr full-array L-band observation, 26 from the 1\,hr full-array S-band observation and 46 from the 2\,hr sub-array UHF observation. These bursts were all detected by the TUSE real-time search pipeline.
Their times of arrival (TOAs, barycentric and referenced to infinite frequency), detected S/Ns, burst widths and fluences are listed in Table~\ref{tab:bursts}. A subset of the bursts, either very bright ($\text{S/N}>50$) or having multiple sub-components and/or complex time-frequency structure, are shown in Figure~\ref{fig:bursts}.

For the PTUSE data of the 2\,hr sub-array L-band observation we performed a standard single pulse search. We first converted the search-mode data to 8-bit filterbanks using {\sc digifil} from the standard pulsar software package {\sc dspsr} \citep{Straten11}. The IQRM algorithm\footnote{\url{https://github.com/v-morello/iqrm}} \citep{Morello22} was used for RFI mitigation. We conducted a boxcar match-filter search for impulsive transient signals over a DM range between 400\text{--}500\pccm\ using {\sc heimdall}. All candidates with $\text{S/N}>8$ were saved for manual inspection. This resulted in the detection of 108 bursts from the 2\,hr sub-array L-band observation, 12 of which were also detected in the simultaneous UHF observation, as indicated in Table~\ref{tab:bursts}. 
However, we used the PTUSE data only for burst rate and spectral extent calculations but not burst morphology or polarisaiton due to issues with the data that are still being investigated. These issues did not affect the FBFUSE/TUSE data or the triggered voltage buffer data.

In summary, we detected 249 bursts in total in our follow-up of \FRB, including 46 in the UHF, 177 in the L-band (including 69 from the full-array observation and 108 from the sub-array observation) and 26 in the S-band. Given 12 bursts were detected simultaneously by the UHF sub-array and L-band sub-array, this reduced to 237 unique bursts. Their TOAs and properties are listed in Table~\ref{tab:bursts}.

\begin{table*}
\begin{threeparttable}
\centering
\resizebox{1.8\columnwidth}{!}{\hspace{-0cm}\begin{tabular}
{l c r r c c c c c c}
\hline
 Burst\tnote{a} & TOA\tnote{b} & $\text{S/N}_\text{det}$\tnote{c} & Burst width\tnote{d} & Fluence\tnote{e} & DM\tnote{f} & Trigger\tnote{g} & RM\tnote{h} & $L/I$\tnote{h} & $|V|/I$\tnote{h} \\
  & (MJD)&  & (ms) & (Jy ms) & (pc cm$^{-3}$) & & ($\text{rad}\,\text{m}^{-2}$) & & \\
\hline
 L1 & 60487.01405828 & 16.9 & 64.8 & 2.32(21) & & N &  &  &  \\
 L2 & 60487.01520363 & 26.8 & 13.6 & 2.70(12) & 467.50(69) & Y & -195(1) & 0.88(11) & 0.12(5) \\
 L3 & 60487.01520510 & 16.6 & 7.0 & 1.20(8) & &N &  &  &  \\
 L4 & 60487.01574760 & 44.9 & 17.0 & 2.45(14) & 463.47(45) & Y & -190(1) & 0.95(9) & 0.13(5) \\
 L5 & 60487.01593774 & 16.2 & 21.3 & 2.07(19) & &N &  &  &  \\
 L6 & 60487.01593947 & 14.3 & 10.9 & 1.31(13) & &N &  &  &  \\
 -- & -- & -- & -- & -- & -- & -- & -- & -- & -- \\
 \hline
\end{tabular}}
\caption{Properties of the repeat bursts detected from \FRB\ with MeerKAT (full table available online).\\
a: Label of each burst with "U" indicating detection at UHF, "L" at L-band and "S" at S-band. The highlighted bursts ($\ast$) were detected simultaneously at UHF and L-band (see Section~\ref{sec:detection}).\\
b: Time of arrival in Barycentric Dynamical Time referenced to infinite frequency.\\
c: Reported S/N by the real-time search pipeline.\\
d: Boxcar equivalent burst width.\\
e: Fluence of each burst estimated by the radiometer equation and the associated error (see Section~\ref{sec:fluence}).\\
f: DM obtained by maximising the coherent power across the bandwidth.\\
g: "Y" and "N" indicate whether the burst triggered the voltage buffer dump or not (see Section~\ref{sec:voltage}).\\
h: Faraday rotation measure and linear and circular polarisation fraction measured from the voltage data (see Section~\ref{sec:polarimetry}).}
\label{tab:bursts}
\end{threeparttable}
\end{table*}

\subsection{Voltage data}\label{sec:voltage}

We saved the voltage data of all triggered bursts along with the gain solutions for all available frequency channels. We can use these data to localise the FRB source. First we correlated the voltage data to create visibilities after applying the gain solutions. Then images were made using {\sc wsclean} \citep{Offringa14, Offringa17} around the time of the burst detection with an integration time down to 0.6\,ms at S-band. Any bright $\sim$ms bursts can be identified as transient sources in these images. See \citet{Rajwade24} for more details on the imaging method.

With the imaging localisation of the FRB source, we can coherently beamform the voltage data to create a high time-resolution, full polarisation time-frequency data product for each burst. We used the {\sc dspsr} package \citep{Straten11} to write these beamformed data into an archive format that can be processed using tools from {\sc psrchive} \citep{Hotan04}, including {\sc pazi} for removing RFI and {\sc pam} for transforming to Stokes parameters. Compared to the real-time data products from the TUSE backend, these archive data have a much higher frequency and time resolution suitable for studying the dynamic pulse structure of the bursts. See \citet{Rajwade24} for more details on the beamforming process.

\renewcommand{\thefigure}{\arabic{figure} (Continued.)}
\addtocounter{figure}{-1}

\begin{figure*}
\centering
\includegraphics[width=.85\textwidth]{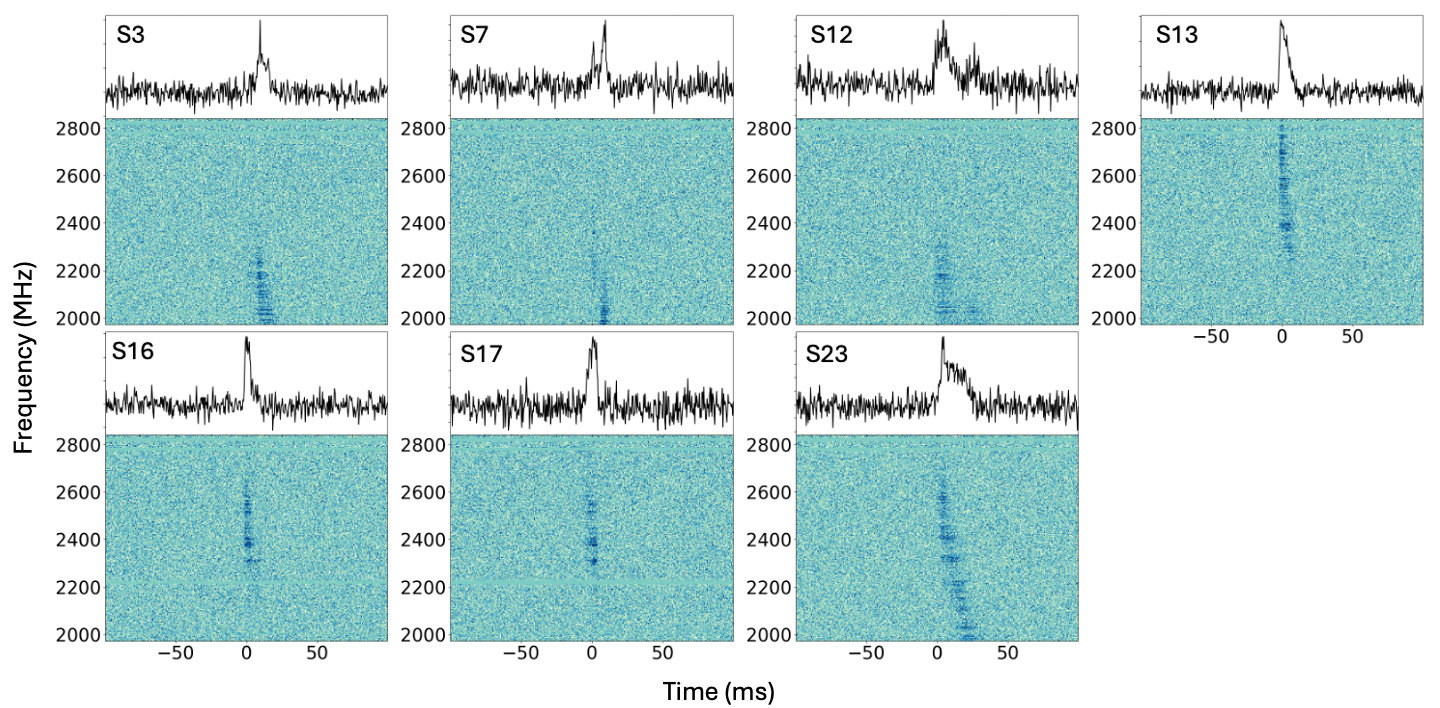}
\caption{}
\end{figure*}

\renewcommand{\thefigure}{\arabic{figure} (Continued.)}
\addtocounter{figure}{-1}

\begin{figure*}
\centering
\includegraphics[width=.85\textwidth]{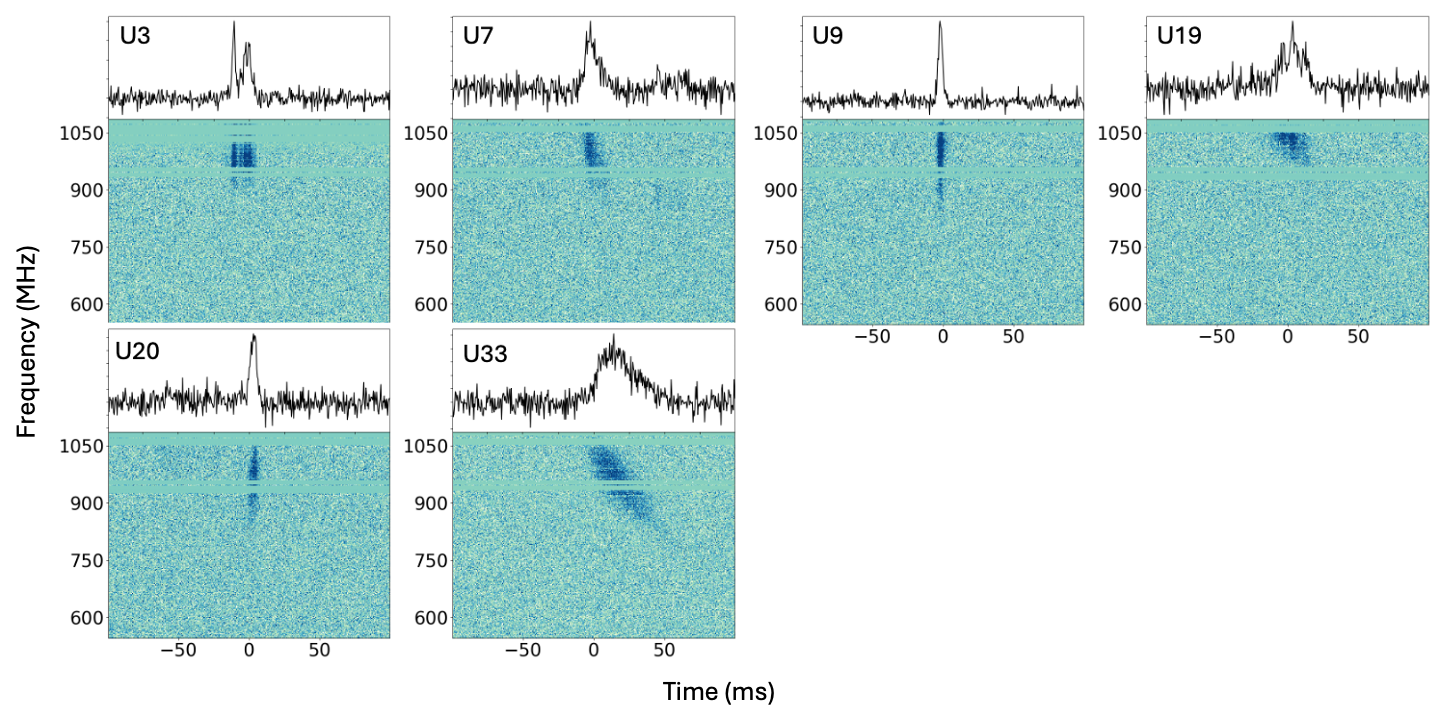}
\caption{}
\end{figure*}

\renewcommand{\thefigure}{\arabic{figure}}

\subsection{Optical observations}

In order to search for transient optical counterparts of \FRB, we conducted contemporaneous optical observations on 2024 June 28 when MeerKAT observed \FRB\ simultaneously in the UHF and L-band. We used MeerLICHT, a fully robotic, wide-field optical telescope with a 0.65\,m primary mirror and a $2.7\,\text{deg}^2$ field of view \citep{Groot24}, to target the field containing the coordinates of \FRB. The spatial resolution of MeerLICHT is approximately $2''$. The observations started with 300\,s exposures of the field in the hours before MeerKAT was due to observe, allowing us to obtain deep images which could be co-added later and used as reference images of the field. Then 15\,s MeerLICHT observations were recorded for several hours, overlapping with the MeerKAT observations. The approximate MeerLICHT limiting magnitudes for exposures of this length are $\sim19.4$ for the $q$-band, $\sim18.3$ for the $i$-band, and $\sim17.6$ for the $u$-band. Note that there is overhead time of approximately $15-20$\,s between exposures for filter change, re-pointing of the telescope, and image readout. All MeerLICHT exposures cycled through the $u$, $q$, and $i$ filters. More details about MeerLICHT's observing strategy can be found in \citet{Hanmer25}.


\section{Results}\label{sec:results}

\subsection{DM estimation}\label{sec:DM}

In detecting the \FRB\ bursts, we used the DMs that maximised the peak S/Ns.
However, given the burst signals of \FRB\ exhibit complex morphology and/or sub-components, as shown in Figure~\ref{fig:bursts}, the S/N maximising DMs could blur the dynamic burst structure. We therefore used a structure optimising approach to measure the DM of \FRB. We selected the most complex burst in our sample, U3, detected in the UHF with a S/N of 36.5 and having many sub-components, and ran {\sc DM\_phase}\footnote{\url{https://github.com/danielemichilli/DM_phase}} \citep{Seymour19}, a DM optimisation algorithm that maximises the coherent power across the bandwidth. We dedispersed the burst over a trial DM range of 460--470\pccm\ in steps of 0.1\pccm, and obtained a structure optimising DM of $464.85\pm0.01$\pccm. The uncertainty was calculated through transforming the standard deviation of the coherent power spectrum into a standard deviation in DM. 
This DM value is consistent with the measurement reported by the AstroFlash team, who determined a structure-optimised DM of $464.87\pm0.11$\pccm\ for a complex, extremely bright burst with $\text{S/N}>300$ detected with a single dish from the WSRT.
Since the DM is not expected to evolve over the time span of our observations (9 days), we adopt $\text{DM} = 464.85\pm0.01$\pccm\  for \FRB\ throughout this work.

\begin{figure}
\centering
\includegraphics[width=.5\textwidth]{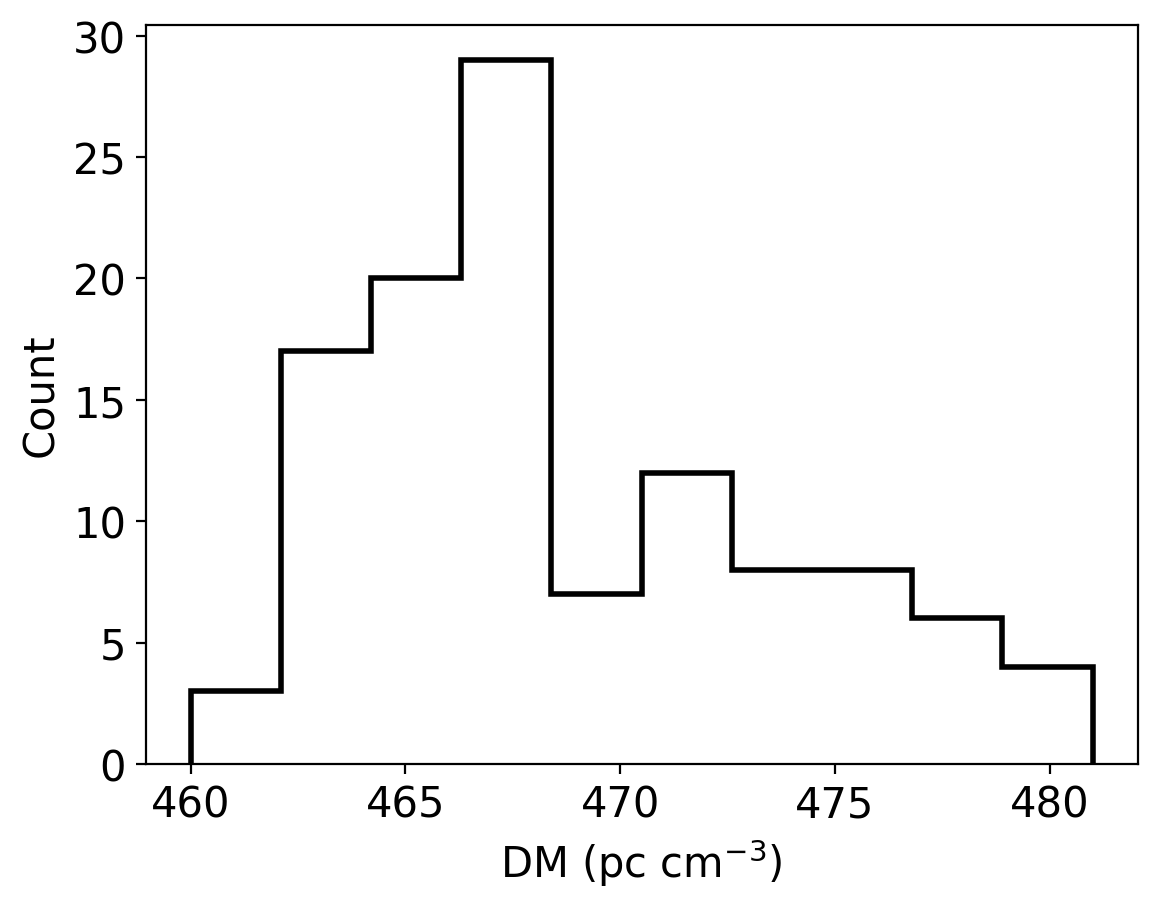}
\caption{Distribution of structure optimising DMs for all repeat bursts detected from \FRB\ with $\text{S/N}>20$.}
\label{fig:DMs}
\end{figure}

We also measured the structure optimising DMs for bright bursts ($\text{S/N}>20$), as given in Table~\ref{tab:bursts}. Figure~\ref{fig:DMs} shows the distribution of the measured DM values. We can see most of the bursts show DMs in the range of $466\text{--}468\,\text{pc}\,\text{cm}^{-3}$. This is consistent with our observation of many under-dedispersed bursts in Figure~\ref{fig:bursts}. However, as the DM is not expected to evolve on timescales of minutes, the observed discrepancy in DM is likely to be an intrinsic property of the bursts, such as different frequency drifts as observed between sub-components in bursts L20, L51 and L56.

\subsection{Imaging and localisation}\label{sec:imaging}

\begin{figure*}
\centering
\includegraphics[width=\textwidth]{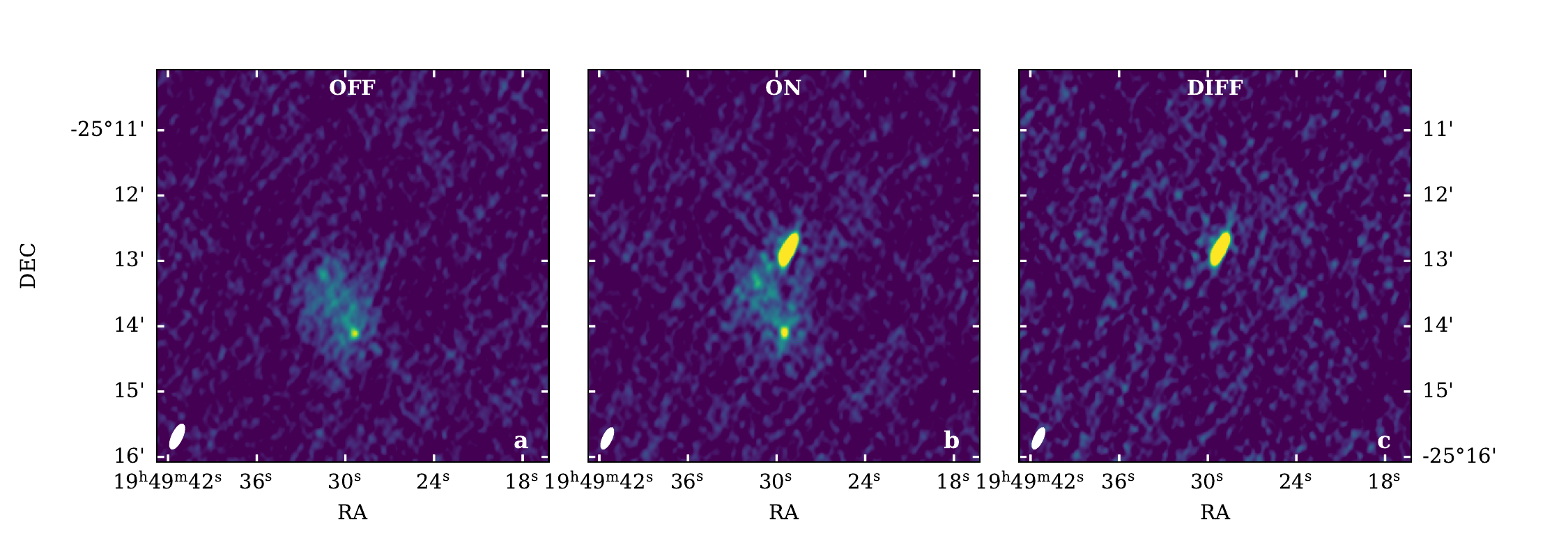}
\caption{Images of the position of \FRB\ before the burst detection (a, off), integrated over the duration of burst L20 (b, on), and the difference between the two (c, diff).
The images have a synthesised beam size of $24.6"\times9.1"$, shown in white on the bottom left corners. The extended source visible on the on and off images is an AGN located near the FRB position.}
\label{fig:imaging}
\end{figure*}

We imaged and localised the FRB following the method described in \cite{Rajwade24}, using a bright L-band burst, L20. While the higher frequencies in the S-band can achieve a better accuracy of the source within the image, fewer radio sources are visible within the field of view, and hence the astrometric error is larger. Additionally, the S-band bursts are generally fainter than the L-band bursts, justifying our choice of an L-band burst for the localisation.
We used the voltage data to create one "on" image where the burst is visible, and one "off" image before the burst occurs, as shown in Figure~\ref{fig:imaging}. We also created a difference image between the on and off images. We used {\sc wsclean} \citep{Offringa14, Offringa17} to reconstruct and deconvolve the radio images from the visibility data using a pixel scale of 1\,arcsec and size of $4096\times4096$ pixels. We identified a transient source in the on-burst image, as shown in Figure~\ref{fig:imaging}. This source appears only at the time of the burst detection, and since there are no other transient sources in the image, we identify this as the FRB location.
Also, the coherent beam where the FRB was detected has a position of RA=19:49:28.79, Dec=-25:12:54.3, which is highly consistent with the identified transient location.
This is further confirmed by the detection of transient sources at the same position using the voltage data of other bursts. Since an active galactic nucleus (AGN) NVSS J194930-251347 \citep{Condon98} located near the FRB position ($\sim1$\,arcmin) was very bright in the L-band images, we used the difference image to better constrain the source position.

After locating the burst in the on and difference images, we performed an astrometric correction. To increase the source count that can be used as an astrometric reference, we created an image using the full 300\,ms of voltage data. The astrometric correction should be the same between the full integration and the on images. 
First, we identified the position of all sources in the images with the Python Blob Detector and Source Finder\footnote{\url{https://www.astron.nl/citt/pybdsf/}} ({\sc PyBDSF}) algorithm. 
Next we selected a reference catalogue to use for the astrometric correction. 
The Rapid ASKAP Continuum Survey mid (RACS-mid) catalogue \citep{McConnell20} contains enough sources within the field of view for the astrometric correction, but given the 1"--2" systematic uncertainty on its source positions, we corrected the astrometry of RACS-mid within a 3\,deg radius using the Radio Fundamental Catalogue \citep[RFC;][]{petrov_radio_2025} as a reference. 
We then selected all the point-like sources with positional errors <0.5" in both RA and Dec from RACS-mid and the MeerKAT full integration image contained within the half-power beam width of the centre \citep[1.1\,deg in the L-band;][]{de_villiers_meerkat_2023}. We identified 8 sources matching between RACS and the MeerKAT image, with separations ranging from 0.3 to 2.4\,arcsec.
These matched sources were used to solve for a transformation matrix to shift and rotate the MeerKAT sources to match the RACS source positions\footnote{The code for performing the astrometric correction can be found on GitHub: \url{https://github.com/AstroLaura/MeerKAT_Source_Matching}}. We then applied the same transformation matrix to the sources in the on and difference images.
Finally, we found the best source position to be RA=19:49:29.21, Dec=-25:12:49.64. The uncertainty consists of three components: the source fitting error given by {\sc PyBDSF} (0.07" in RA and 0.14" Dec), the absolute astrometric uncertainty from the RACS positions (0.17" in both RA and Dec) and the median offset of the positions after the astrometric correction (0.85"). We added these uncertainties in quadrature and found the total uncertainty to be 0.9" in RA and Dec.


\begin{figure*}
    \centering
    \includegraphics[width=\textwidth]{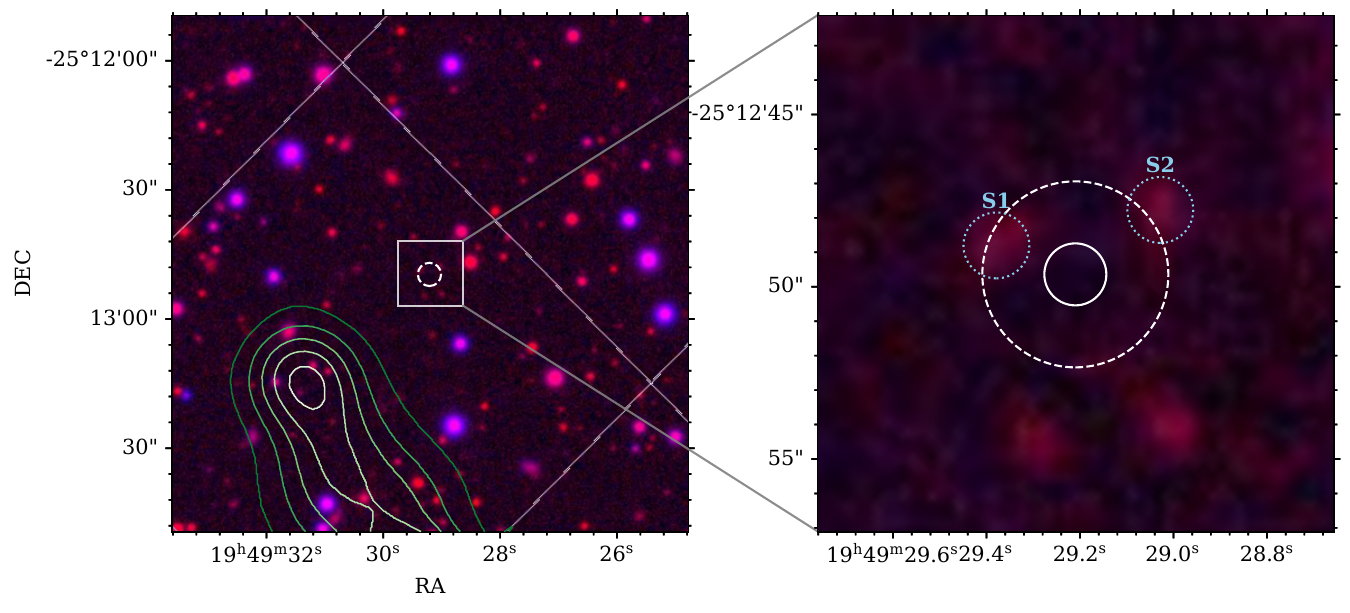}
    \caption{Localisation of \FRB\ with the DESI archival images (g, i filters) as background. The left image shows a $120"\times120"$ field of view, with green contours representing the extended radio source near the FRB location, while the right image shows a zoomed version $15"\times15"$ field of view. The white solid circle gives the $1\sigma$ localisation error and the white dashed circle the $3\sigma$ error. Sources S1 and S2 are the two candidate host galaxies. }
    \label{fig:optical}
\end{figure*}

In order to identify the host galaxy where this FRB might have originated, we searched archival optical images at the FRB location. The deepest public catalogue at these coordinates is the DESI-DR10 \citep{Dey19}, where we identified two sources, S1 and S2, at 2.4" and 3.0" respectively from the FRB position, as shown in Figure~\ref{fig:optical}. 
Because both these sources fall outside the $1\sigma$ uncertainties on the FRB location, fainter host galaxy candidates might be located closer to the FRB localisation. Deep optical observations will be presented in an upcoming publication.

\subsection{Burst morphology}\label{sec:morphology}

We observe a wide range of burst morphologies in the sample of repeat bursts detected from \FRB, as shown in Figure~\ref{fig:bursts}. 
In this work, we distinguish independent bursts using a criteria of their separation being $\gtrsim80$\,ms (larger than the widest pulse in our sample).
While most of the bursts show single-peaked pulse profiles, some consist of many sub-components, e.g. L17, L27 and L47. These complex bursts mostly occur in the L-band. 
The multiple burst components of L27 drift downwards in frequency, as typically observed in repeating FRBs \citep{Pleunis21b}
However, many bursts show more complex spectro-temporal structures. In L47, L51 and L56, the intra-burst drift rate varies between sub-components. 
Note that the intra-burst drift is defined as the drift of the emission within a sub-burst, and the rate can be calculated as the derivative of emission frequency with respect to time.
In addition, L47 exhibits both downward and upward inter-burst drifts. Such complex drifting phenomenologies have been observed in both repeating and non-repeating FRBs \citep{Faber24, Curtin24}. Here, for simplicity, we select only the single-peaked bursts to measure the drift rate. We used the standard technique, i.e. computing the 2D autocorrelation function (ACF) of the dynamic spectrum and fitting a 2D Gaussian function to calculate the tilt in the semi-major axis of the ACF. This resulted in a drift rate between $-61.8\,\text{MHz}\,\text{ms}^{-1}$ (L34) and $55.4\,\text{MHz}\,\text{ms}^{-1}$ (L55) with an RMS value of $33.5\,\text{MHz}\,\text{ms}^{-1}$
for the L-band bursts. We performed a similar analysis on the bright bursts in the UHF and S-band, and measured a drift rate of $238.4\,\text{MHz}\,\text{ms}^{-1}$ for U9, $-5.6\,\text{MHz}\,\text{ms}^{-1}$ for U33 and $-29.2\,\text{MHz}\,\text{ms}^{-1}$ for S23. The uncertainty is not well constrained with the direct Gaussian fitting approach, so it is not included here.

\begin{figure*}
\centering
\includegraphics[width=\textwidth]{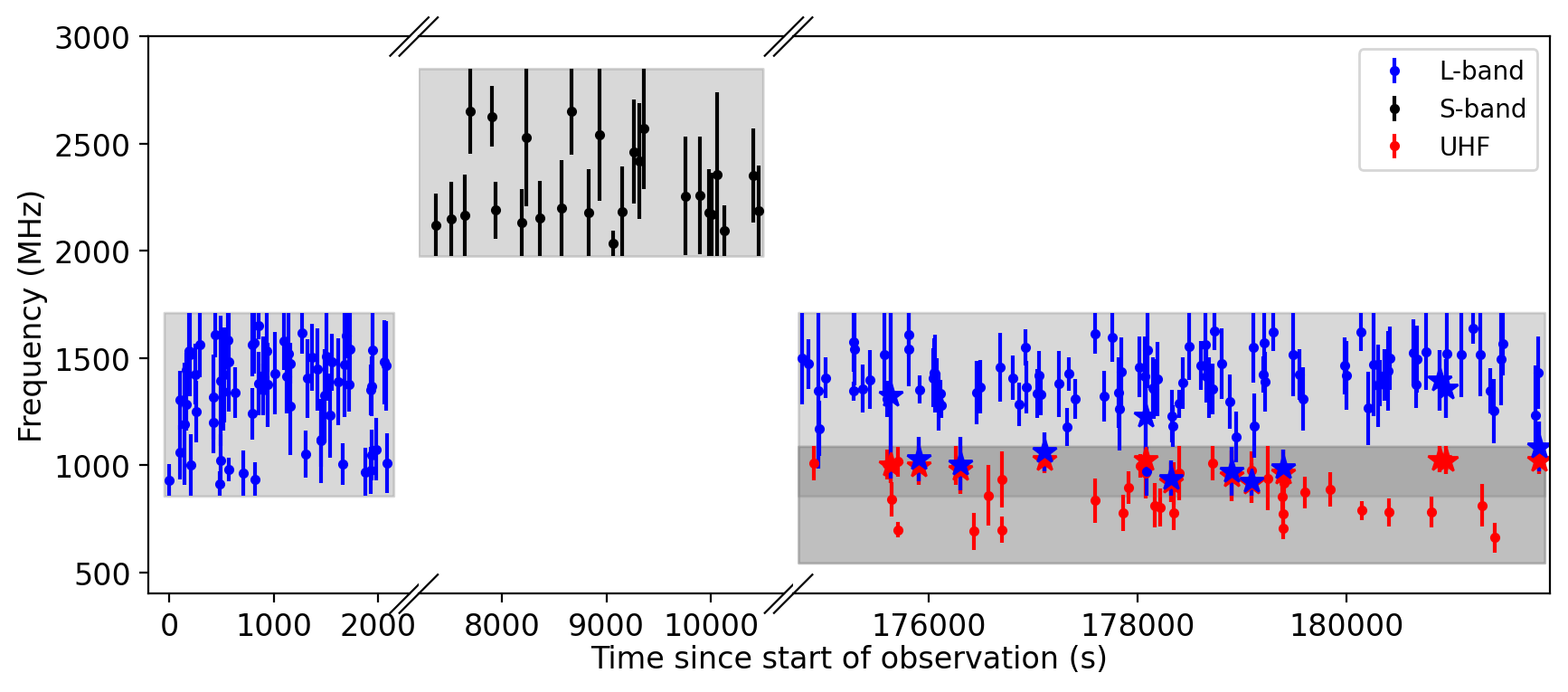}
\caption{The emission frequency of each MeerKAT-detected burst from \FRB\ is plotted against the burst arrival time. Each point represents the central emission frequency of a burst with the error bar corresponding to the spectral extent. The S-band and L-band observations on 2024 June 26 were separated by a $\sim90$\,min gap, and the simultaneous UHF and L-band observations were $\sim2$\,day later (see Section~\ref{sec:observation}). The shaded regions indicate the bandwidths of the UHF, L-band and S-band observations. The stars highlight the 12 bursts detected simultaneously at UHF and L-band (also highlighted in Table~\ref{tab:bursts}).
The right panel shows the distribution of the central observed frequency in the UHF (red), L-band (blue) and S-band (black).}
\label{fig:spec_extent}
\end{figure*}

The MeerKAT-detected repeat bursts have a wide range of spectral extents. Most of them are band-limited, while a few span almost the full extent of the observing band, e.g. L42 and S23. In order to quantify the spectral extent, we fitted a Gaussian function to the spectrum of the on-pulse region of each burst. The full width at 10\% of the maximum was adopted as the spectral extent. Figure~\ref{fig:spec_extent} shows the spectral extent of the \FRB\ bursts as a function of burst arrival time. The bandwidths of the UHF, L-band and S-band are indicated by the shaded regions. There is a $\sim90$\,min gap between the L-band and S-band observations on 2024 June 26 and a $\sim2$\,day gap between the first and second follow-up (see Section~\ref{sec:observation}). 
During the second follow-up observation, there were 12 bursts simultaneously detected at UHF and L-band, as highlighted in Figure~\ref{fig:spec_extent} (also highlighted in Table~\ref{tab:bursts}). As these bursts show distinct sub-components with different spectral extents, our spectral measurements on different sub-components of the same burst at UHF and L-band are also different, especially for bursts L82 (U3), L118 (U19), L165 (U42) and L166 (U43). Meanwhile, some bursts have most of their emission in the overlapping region of the UHF and L-band and thus have similar spectral extent measurements in these two bands, e.g. L85 (U7), L94 (U9), L124 (U23) and L138 (U28).
The spectral extents of the repeat bursts range from 69.6\,MHz to 296.1\,MHz in the UHF, 89.4\,MHz to 774.7\,MHz in the L-band and 117.5\,MHz to 764.1\,MHz in the S-band with an average of 169.4\,MHz, 287.2\,MHz and 423.5\,MHz, respectively. This corresponds to a fractional bandwidth of 0.31, 0.34 and 0.48 in the three bands. Note that the signal spectral envelope extends to the top or bottom of the bandwidth in many bursts, as can be seen in Figure~\ref{fig:spec_extent}, and our spectral extent measurements should be taken as lower limits for these bursts. The right panel in Figure~\ref{fig:spec_extent} shows the distribution of the central emission frequency of the repeat bursts detected in the UHF (red), L-band (blue) and S-band (black). As can be seen, the burst rate in the L-band is much higher than in the other two bands (see Section~\ref{sec:burst_rate}). This may suggest a preferential emission frequency around $\sim1.4$\,GHz for \FRB. 
The two peaks in the central frequency distributions of the UHF and S-band bursts indicate clustering of bursts at the top of the UHF and bottom of the S-band. This also suggests a preferential frequency in between these two bands, i.e. $\sim1.1\,\text{--}\,1.9$\,GHz.

\begin{figure*}
\centering
\includegraphics[width=\textwidth]{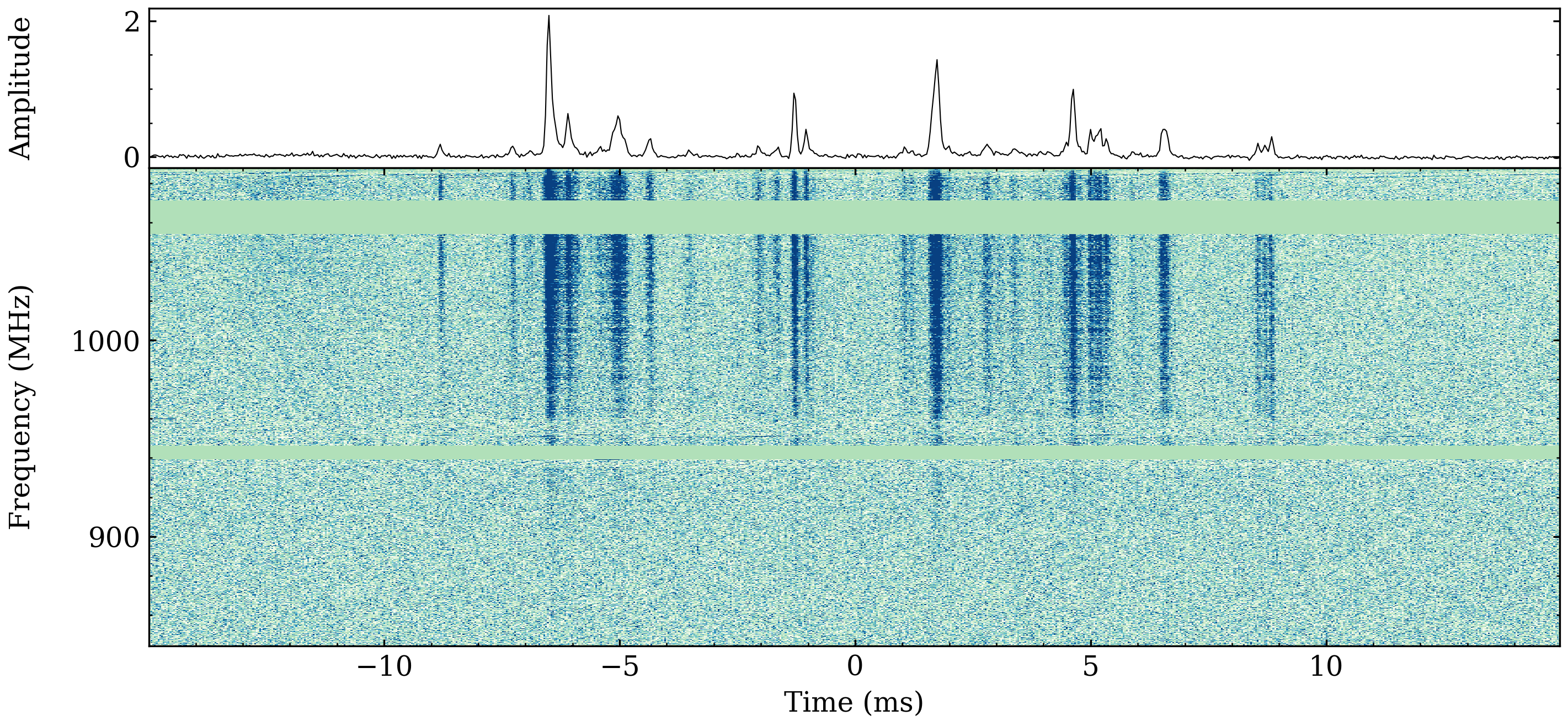}
\caption{Dynamic spectrum of U3, the most complex burst detected from \FRB, plotted using the triggered voltage data. We coherently dedispersed the data to $\text{DM}=464.85$\pccm. The frequency and time resolution are 0.27\,MHz and $32.86\,\mu$s, respectively, to display the microstructure in the burst. Channels are manually masked to remove RFI.}
\label{fig:U3}
\end{figure*}

\begin{figure}
\centering
\includegraphics[width=.5\textwidth]{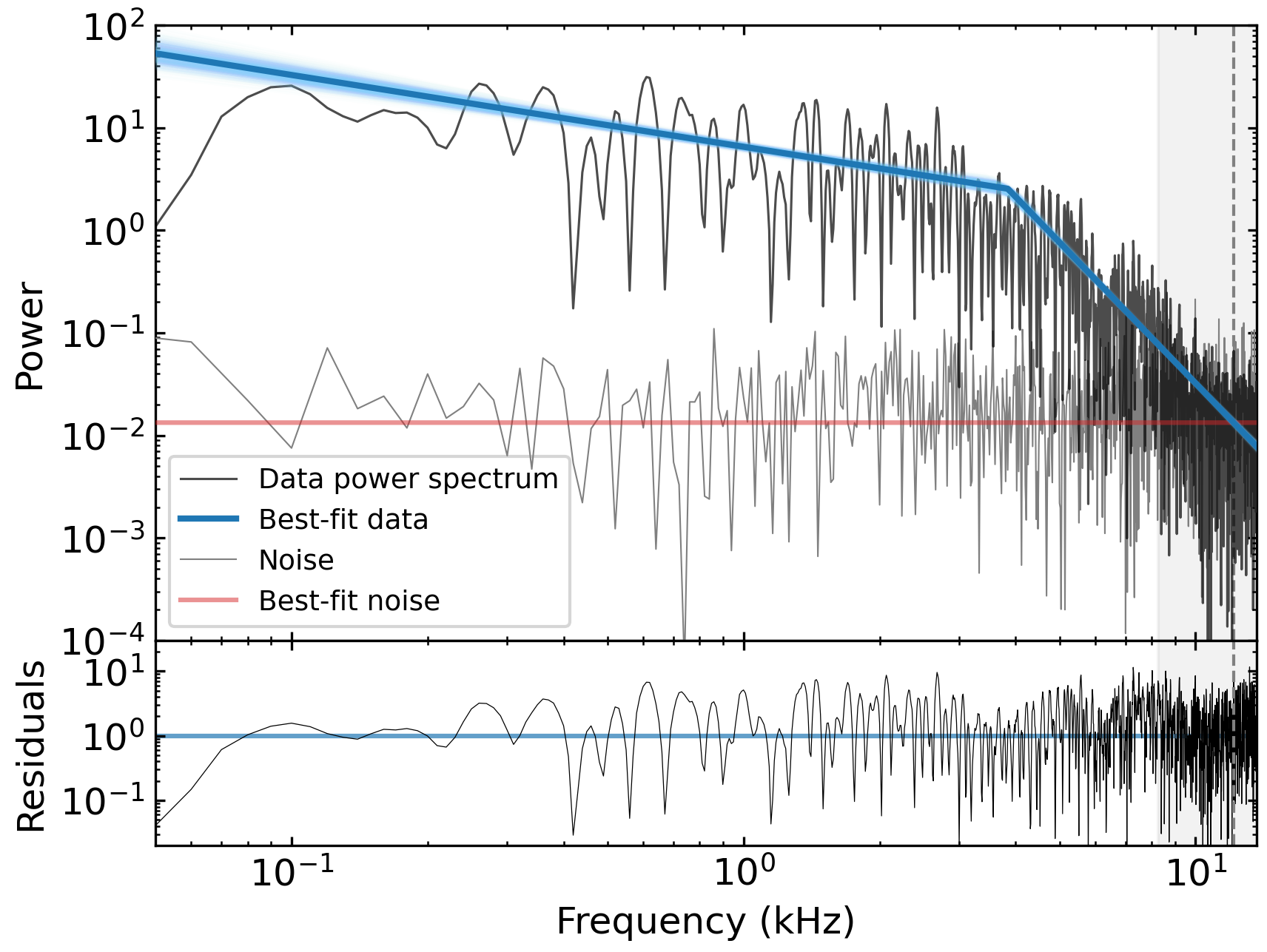}
\caption{Power spectra of the temporal profile of burst U3 and the off-burst noise shown in solid black and grey lines, respectively. We fitted the burst power spectrum with a broken power law, as shown in blue. The solid line represents the best-fit model, while the shaded region corresponds to a random subset of posterior samples after burn-in, illustrating the uncertainty in the fit.
The red line indicates the noise level, while the dotted vertical line indicates the crossover frequency where the power law intersects the noise spectrum. The bottom panel shows the residuals of the fit.}
\label{fig:fft_U3}
\end{figure}

The most complex burst in our sample (excluding the bursts detected in the sub-array L-band observation) is U3. Utilising the triggered voltage data, we replot this burst with a much higher time resolution of $32.86\,\mu$s in Figure~\ref{fig:U3}. As can be seen, U3 consists of a large number of microstructures. In order to search for quasi-periodicity in these microstructures, we computed the power spectrum of the temporal profile of U3, as shown in Figure~\ref{fig:fft_U3}. 
No obvious peaks are observed in the power spectrum that could correspond to a period, but it appears to show a turnover at a certain frequency, hence we fitted it to a broken power law:
\begin{equation}
    f(x) =  
    \begin{cases}
      A x^{-\alpha_1} + C & \text{ if } x\leq x_0\\
      B x^{-\alpha_2} + C & \text{ otherwise,}
    \end{cases}    
\end{equation}

\noindent where $A$ and $B$ are constants, with $B=A\times x_0^{\alpha_2-\alpha_1}$ for continuity, $\alpha_1$ and $\alpha_2$ are the indices before and after the break, and $x_0$ is the break frequency. 
We modelled the power spectrum using a Markov Chain Monte Carlo (MCMC) method implemented with the \texttt{emcee} Python package \citep{foreman-mackey_emcee_2013}. We defined uniform log-priors within manually selected limits, enforcing the constraint $\alpha_1<\alpha_2$. The log-posterior distribution was computed using Bayes’ theorem, and the MCMC sampling was performed on the resulting distribution. We initialised the sampler with 16 walkers and ran it for 10,000 steps, discarding the first 7,000 samples as burn-in. The final parameter estimates obtained from the posterior distribution are the following: $A= 6.5\pm0.5$, $\alpha_1=0.7\pm0.1$, $\alpha_2=4.7\pm0.2$, $x_0=3.8\pm0.4$\,kHz, and $\log_{10}C= -4.2^{+1.2}_{ -0.8}$.
To determine if there are any significant outliers that could correspond to a (quasi-)periodicity in the pulse profile, we first computed the residuals defined as $R(x)=2P(x)/M(x)$, where $P(x)$ is the power spectrum of the data and $M(x)$ is the best-fit model, as shown on the bottom panel of Fig.~\ref{fig:fft_U3}. We compared the residuals to a $\chi^2$ distribution with two degrees of freedom by applying a Kolmogorov-Smirnov (KS) test \citep{timmer_generating_1995, vaughan_simple_2005}, and found a p-value of 0.53. This means that we cannot reject the null hypothesis that the residuals are drawn from a $\chi^2$ distribution. This implies that the pulse profile of U3 shows no evidence of periodicity. However, because the broken power law might dominate over any quasi-periodicity, and any structure below $32.86\,\mu$s is unresolved, a more comprehensive analysis would be required to rule out periodicities in those regimes, which is out of the scope of this work.
Nevertheless, the break frequency $x_0=3.8\pm0.4$\,kHz indicates that the burst structure follows a typical timescale of $1/x_0=0.26$\,ms.
The resulting value of $\alpha_1$ is comparable to those measured for other active repeaters \citep{Nimmo21, Nimmo22, Hewitt23, pastor-marazuela_fast_2023}. 
To explore the shortest timescale in this burst, we measured the critical frequency beyond which the burst power spectrum drops to below the off-burst noise. The result is shown in dotted black line in Figure~\ref{fig:fft_U3} with the error (shaded region) determined by the uncertainties in the power-law fit. This frequency corresponds to a timescale of $\sim80\,\mu$s that can be marginally resolved by the time resolution. We cannot further constrain the microstructures with shorter integrations due to S/N limitations.


Some of the MeerKAT-detected repeat bursts show brightness variations in the spectrum, probably due to diffractive scintillation in the Galactic interstellar medium. Detailed study of the scintillation bandwidth and scattering timescale is beyond the scope of this paper and will be undertaken in future work.

\subsection{Fluences}\label{sec:fluence}

We used the modified single-pulse radiometer equation \citep{Dewey85} from \cite{Jankowski23} to estimate the fluence of each burst:

\begin{equation}
    S_\text{peak} (\text{S/N}, W_\text{eq}, \vec{a}) = \text{S/N}\,\beta\,\eta_\text{b}\,\frac{\text{SEFD}}{\sqrt{b_\text{eff}N_\text{p}W_\text{eq}}}\,a^{-1}_\text{CB}\,a^{-1}_\text{IB},
    \label{eq:radiometer}
\end{equation}

\noindent where $S_\text{peak}$ is the peak flux density, $\vec{a}=(a_\text{CB}, a_\text{IB})$ are the attenuation factors of the detection CB and incoherent beam (IB), $\beta\approx1$ is the digitization loss factor, $\eta_\text{b}\approx1$ is the beamforming efficiency, $\text{SEFD}$ is the system equivalent flux density of the MeerKAT array\footnote{See the online MeerKAT technical documentation: \url{https://science.ska.ac.za/meerkat}}, $b_\text{eff}$ is the effective bandwidth in Hz, $N_\text{p}=2$ is the number of polarisations, and $W_\text{eq}$ is the observed equivalent boxcar pulse width in seconds. In the above equation, three parameters, $\text{SEFD}$, $a_\text{CB}$ and $a_\text{IB}$, are frequency dependent and well characterised for the MeerKAT array. The SEFD at L-band can be fit by a polynomial function \citep{Geyer21}
\begin{equation}
    \text{SEFD} = 5.71\times10^{-7}\,\nu^3 - 1.90\times10^{-3}\,\nu^2 + 1.90\,\nu -113,
\end{equation}

\noindent where $\nu$ is the observing frequency in MHz and $\text{SEFD}$ in Jy. Similarly, the SEFD at UHF can be expressed as
\begin{equation}
    \text{SEFD} = 2.30\times10^{-6}\,\nu^3 - 4.69\times10^{-3}\,\nu^2 + 2.52\,\nu + 286,
\end{equation}

\noindent and at S-band as
\begin{equation}
    \text{SEFD} = 2.40\times10^{-7}\,\nu^3 - 1.50\times10^{-3}\,\nu^2 + 3.05\,\nu - 1525.95.
\end{equation}

\noindent Note that the above $\text{SEFD}$ is for a single MeerKAT dish, and we need to divide that by the number of antennas used for the observation. The beam correction $a_\text{IB}$ and $a_\text{CB}$ was calculated using the MeerKAT primary beam model ({\sc katbeam}\footnote{\url{https://github.com/ska-sa/katbeam}}) and coherent beam model ({\sc mosaic}\footnote{\url{https://github.com/wchenastro/Mosaic}}; \citealt{Chen21}) for the \FRB\ position obtained in Section~\ref{sec:imaging}.

In estimating the fluences of the repeat bursts of \FRB, considering their limited spectral extents (see Section~\ref{sec:morphology}), we split the full bandwidth into 8 sub-bands. We measured the S/N of the burst dedispersed to the optimal DM (see Section~\ref{sec:DM}) in each sub-band and converted it to a flux density using Eq.~\ref{eq:radiometer} and fluence $F=S_\text{peak}\,W_\text{eq}$. Note that $\text{SEFD}$, $a_\text{CB}$ and $a_\text{IB}$ are frequency dependent and thus have different values in different sub-bands, and $b_\text{eff}$ is the effective bandwidth of the sub-band excluding flagged channels. The final fluence of each burst summed from all the sub-bands is listed in Table~\ref{tab:bursts}, with the uncertainty corresponding to $1\sigma$ radiometer noise.

\begin{figure}
\centering
\includegraphics[width=.5\textwidth]{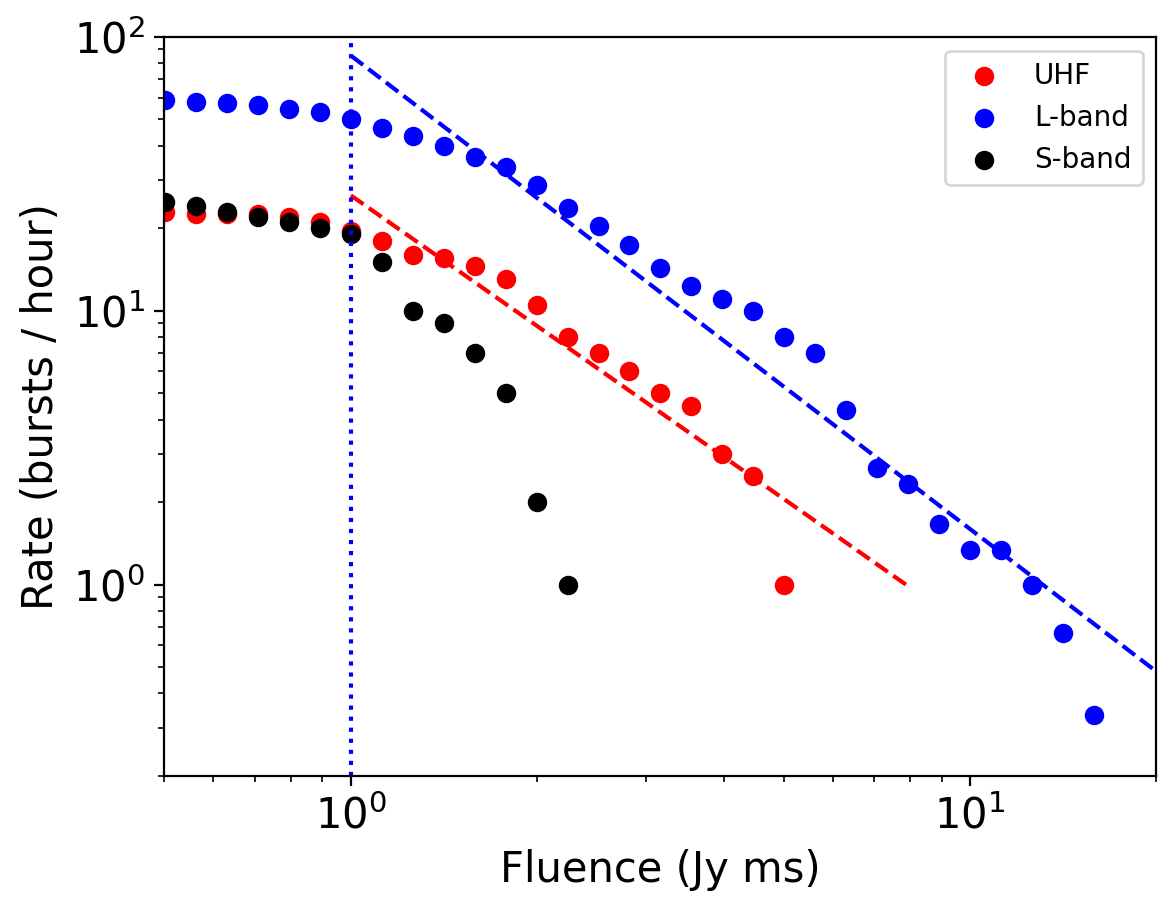}
\caption{Cumulative burst rate function of the MeerKAT detected bursts at UHF, L-band and S-band. The vertical dotted line marks the fluence completeness limit of MeerKAT observations, and the dashed lines in different colors show the best-fitting power law for the bursts above the completeness level.}
\label{fig:fluence}
\end{figure}

Figure~\ref{fig:fluence} shows the cumulative distribution of the MeerKAT detected burst rate at UHF, L-band and S-band, respectively, above a given fluence. There is a break at $\sim1$\,Jy\,ms in the distribution of the L-band bursts, which can be attributed to the completeness limit of MeerKAT observations. Previous studies of the MeerTRAP survey performance and completeness found the fluence completeness limit to be 0.7\,Jy\,ms \citep{Jankowski23}, consistent with the value observed here. 
Note that there is a slight sensitivity difference between the full array and sub-array observations due to their different numbers of antennas (40 vs. 32; see Section~\ref{sec:observation}).
Fitting the cumulative distribution above the completeness limit with a function of the form $R (>F)\propto F^\gamma$ yielded a power-law index of $\gamma=-1.6\pm0.1$ and $\gamma=-1.7\pm0.1$ at UHF and L-band, respectively. We do not provide fitting results for the small sample of S-band bursts. The two power-law indices are consistent with each other within a $1\sigma$ error. They are comparable to the values measured for FRB~20121102A at 1.4\,GHz ($-1.8\pm0.3$; \citealt{Gourdji19, Aggarwal21}), FRB~20180916B at 600\,MHz ($-1.3\pm0.3\pm0.1$; \citealt{CHIME20b}) and FRB~20201124A at 650\,MHz ($-1.2\pm0.2$; \citealt{Marthi22}). We note that setting a different break in fluence would affect the fitting result, e.g., a break at 2\,Jy\,ms would give a power-law index of $\gamma=-2.2\pm0.3$ at UHF and $\gamma=-2.0\pm0.1$ at L-band, steeper than that with the 1\,Jy\,ms break. A more rigorous analysis of the fluence completeness limit is beyond the scope of this paper.

\subsection{Burst rate and arrival times}\label{sec:burst_rate}

We detected 46, 177 and 26 bursts in the MeerKAT UHF (2\,hr), L-band (3\,hr, 1\,hr with the full array and 2\,hr with the sub-array; see Section~\ref{sec:observation}) and S-band (1\,hr) observations, respectively, with their TOAs being listed in Table~\ref{tab:bursts}. 
As the full-array used only 40 out of the 64 dishes, the sub-array is approximately 20\% less sensitive than the full-array. We therefore neglect the sensitivity loss of the sub-array and its influence on the detection rates.
If all bursts above our detection limit, 0.08\,Jy\,ms at UHF, 0.04\,Jy\,ms at L-band and 0.05\,Jy\,ms at S-band (the faintest bursts detected in these three bands), have been detected, we obtain a burst rate of $23\,\text{hr}^{-1}$, $59\,\text{hr}^{-1}$ and $26\,\text{hr}^{-1}$ at UHF, L-band and S-band, respectively. However, given the incomplete fraction of bursts $\lesssim1$\,Jy\,ms at UHF and L-band and the small sample of S-band bursts, these burst rate estimates are likely to be inaccurate. If we only consider the bursts above the completeness limit of $\sim1$\,Jy\,ms at UHF and L-band, the burst rates in these two bands would be $19.5\,\text{hr}^{-1}$ and $50\,\text{hr}^{-1}$, qualifying \FRB\ as a hyperactive repeater. We also notice the \FRB\ source is much more active in the L-band than in the UHF and S-band, as can be seen in Figure~\ref{fig:fluence}. Assuming the same fluence limit for the three bands, we find \FRB\ emits most bursts at 1284\,MHz, which are $3\times$ more than at 816\,MHz and 2405\,MHz. This frequency-dependent activity has also been observed in other repeaters (e.g. \citealt{Pastor21}; \citealt{Houben19, Kumar24c}).

\begin{figure*}
\subfigure[UHF bursts]{
  \label{fig:wait_UHF}
  \includegraphics[width=.49\linewidth]{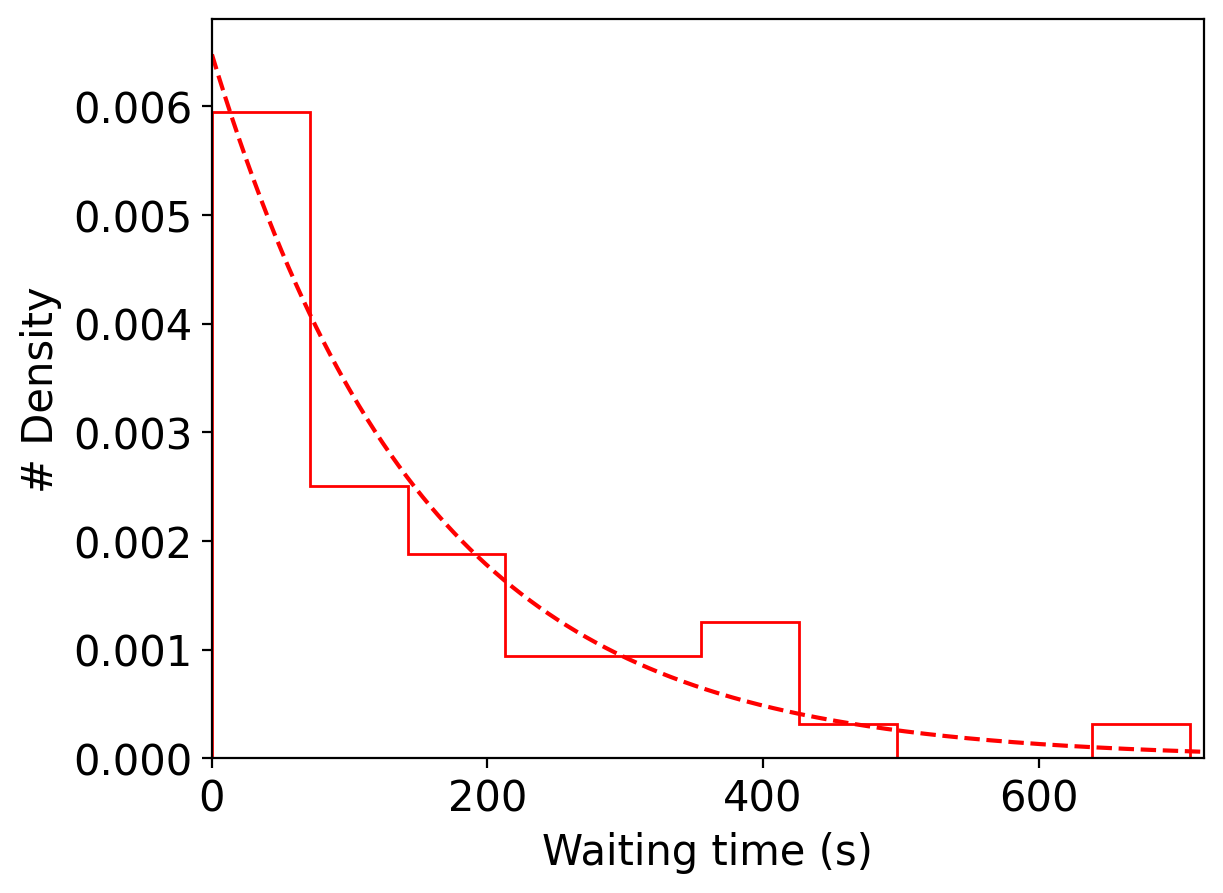}}
\subfigure[L-band bursts]{
  \label{fig:wait_L}
  \includegraphics[width=.49\linewidth]{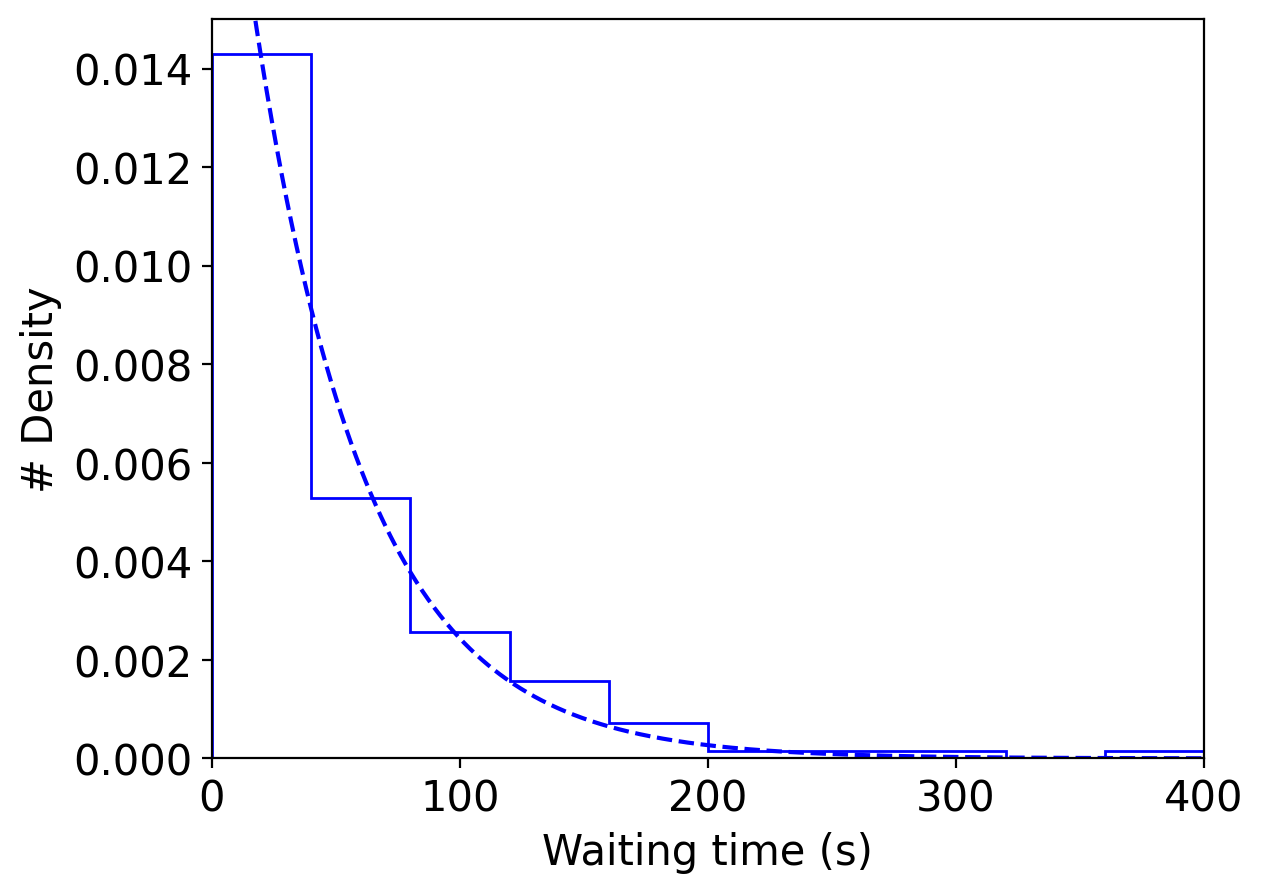}}
\caption{Distribution of the burst waiting time in the MeerKAT UHF (left) and L-band (right) observations. The dashed line shows the Poisson distribution with a constant rate given by the mean waiting time, $\lambda\sim1/154\,\text{s}^{-1}$ in the UHF and $\lambda\sim1/31\,\text{s}^{-1}$ in the L-band, respectively.}
\label{fig:wait_distribution}
\end{figure*}

Figure~\ref{fig:wait_distribution} shows the distribution of waiting times between adjacent bursts detected from \FRB\ in the MeerKAT UHF and L-band. We do not show the distribution for the S-band bursts due to their smaller sample. The mean waiting time between bursts within the 2\,hr UHF and 3\,hr L-band observation is $\sim154$\,sec and $\sim52$\,sec, respectively. If the burst occurrence follows a Poisson process, the waiting time should be exponentially distributed with a probability density function 

\begin{equation}
f(t)=\lambda\exp{(-\lambda t)},
\label{eq:poisson}
\end{equation}

\noindent where $\lambda$ is a rate parameter. 
In order to test whether the burst rate in our observations is Poissonian, we compared the observed distribution of waiting times and the Poisson distribution with a constant rate $\lambda=1/\langle\text{mean waiting time}\rangle$, i.e. $\sim1/154\,\text{s}^{-1}$ in the UHF and $\sim1/52\,\text{s}^{-1}$ in the L-band, as shown in Figure~\ref{fig:wait_distribution}. We found weak evidence for the observed burst waiting times being consistent with the constant-rate Poissonian repetition. The two-sided KS test that measures the maximum distance between the empirical cumulative distribution function (CDF) and the CDF corresponding to Eq.~\ref{eq:poisson} yields a $p$-value of $\approx0.18$ for the UHF bursts and $\approx0.19$ for the L-band bursts. 
Note that this result only applies to the MeerKAT observations during which \FRB\ was extremely active. 
Long-term monitoring of the FRB source is required to reveal any clustering behaviour in the burst repetition during bursting activity, as has been demonstrated for other repeaters (e.g. \citealt{Oppermann18, Lanman22})

\subsection{Polarimetry}\label{sec:polarimetry}

Of the 141 MeerKAT detected repeat bursts from \FRB\ (excluding the bursts detected in the sub-array L-band observation), 40 (11 in the UHF, 20 in the L-band and 9 in the S-band) triggered voltage buffer dumps and thus have polarisation information. In order to study their polarisation properties, we need to correct for the Faraday rotation first. This was done by measuring the RM of each burst using the {\sc rmsynth} tool from the software {\sc psrsalsa}\footnote{\url{https://github.com/weltevrede/psrsalsa}}, a suite of algorithms for statistical analysis of pulsar data \citep{Weltevrede16}. We chose a broad RM search range between $\pm10000\,\text{rad}\,\text{m}^{-2}$ with a step size of $0.1\,\text{rad}\,\text{m}^{-2}$. We find a mean value of $-185\,\text{rad}\,\text{m}^{-2}$, and the measured RMs for all bursts are listed in Table~\ref{tab:bursts}.

\begin{figure*}
\centering
\includegraphics[width=.91\textwidth]{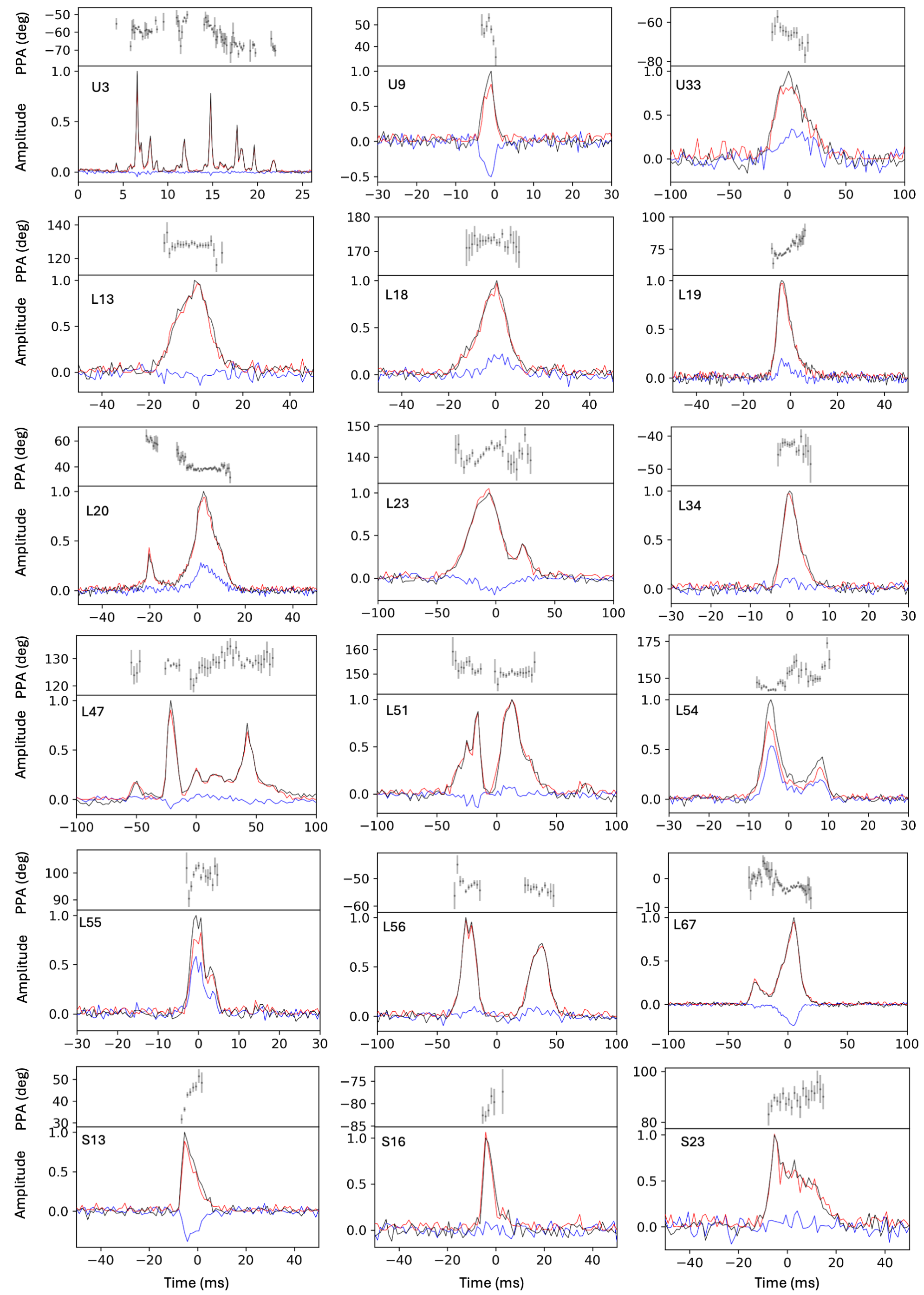}
\caption{Polarisation profiles of a selection of MeerKAT detected bursts from the \FRB\ source. Each panel shows the PPA (top) and the frequency averaged pulse profile (bottom) for total intensity ($I$, black), linear polarisation ($L$, red) and circular polarisation ($V$, blue). The polarisation data are Faraday corrected to the RM values listed in Table~\ref{tab:bursts}.}
\label{fig:pol}
\end{figure*}

After correcting the Stokes spectra of individual bursts, we used {\sc psrsalsa} to measure their polarisation fraction and polarisation position angle (PPA) as a function of time. Specifically, we used {\sc pmod} to remove the baseline of the Stokes parameters and {\sc ppol} to calculate the linear polarisation intensity $L=\sqrt{Q^2+U^2}$ and PPA. We removed the bias in $L$ for each time sample using \citep{Wardle74}:

\[
 L_\text{de-bias} = 
  \begin{cases} 
   L\sqrt{1-\left(\frac{\sigma}{L}\right)^2} & \text{if } 
   L>\sigma \\
   0 & \text{otherwise,}
  \end{cases}
\]

\noindent where $\sigma=\sqrt{(\sigma_Q^2+\sigma_U^2)/2}$ is the off-pulse standard deviation. A $3\sigma$ threshold on the de-biased $L$ was used to establish the significant measurements of PPA. A subset of polarimetric pulse profiles, including 3 at UHF, 12 at L-band and 3 at S-band, along with the significant measurements of PPA, are plotted in Figure~\ref{fig:pol}. We averaged $L/I$ and $|V|/I$ across the pulse profile to obtain the linear and circular polarisation fractions, as shown in Table~\ref{tab:bursts}. Uncertainties are computed from the off-pulse standard deviation in the Stokes parameters based on the principle of error propagation. Note that given the proximity between the \FRB\ source and the center of the primary beam of MeerKAT ($\sim8$\,arcsec; see Section~\ref{sec:imaging}), we expect polarisation leakages to be negligible in our polarisation measurements.

There are a diversity of PPA variations across the pulse profiles in Figure~\ref{fig:pol}, including a nearly constant PPA for most of the bursts, sweeping up for L19 and S13, and sweeping down for U33. Interestingly, the complex burst L20 shows a PPA jump of $\sim20$\,deg in between the two sub-components, while the PPA within each sub-component is constant. These are reminiscent of the PPA variations observed in bursts detected from FRB~20180301A \citep{Luo20} and FRB~20240114A \citep{Tian24b}. We also find most of the MeerKAT-detected repeat bursts from \FRB\ have $\sim100\%$ degree of linear polarisation with no evidence of frequency dependent depolarisation. Meanwhile, most of the bursts show $\sim10\%\text{--}20\%$ degree of circular polarisation with the largest $|V/I|\sim50\%$ being observed in U9, L54 and L55.

\subsection{Optical counterpart search}

Comparing the TOAs of the 142 unique bursts detected on 2024 June 28 with the start and end times of the MeerLICHT exposures, we found 80 bursts with strictly simultaneous optical coverage.
We started by co-adding all of the long exposures we took of the field of \FRB, using MeerLICHT’s co-adding script {\sc buildref.py}\footnote{\url{https://github.com/pmvreeswĳk/BlackBOX/blob/main/buildref.py}}. These reference images were then subtracted from individual MeerLICHT images that contained bursts. The resulting images were further processed using MeerLICHT’s forced photometry package {\sc force\_phot.py}\footnote{\url{https://github.com/pmvreeswĳk/ZOGY/blob/main/force_phot.py}}, which allowed us to extract limiting magnitudes at a $3\sigma$ significance level at the FRB coordinates. These limiting magnitudes were converted to flux limits and then fluence limits, by multiplying each flux limit by the exposure time. The Gaia DR2 astrometric catalogue was used for MeerLICHT astrometric calibration, while low-resolution Gaia DR3 spectra for stars in the field of view were used for photometric calibration \citep{Groot24}. We also created an “on” image per MeerLICHT filter, by co-adding all MeerLICHT images that contained one or more burst TOAs, and ran the image subtraction and forced photometry on these images. All of the photometry results were corrected for Milky Way extinction based on the \citet{Schlafly11} colour excess map and assuming the \citet{Fitzpatrick99} reddening law with the total-to-selective extinction ratio $R_V=3.1$.

We found no optical counterpart to \FRB\ in our simultaneous MeerLICHT observations of individual bursts or the co-added "on" images. The deepest optical fluence limit is $F_\text{opt}<0.76$\,Jy\,ms on a timescale of 15\,s for burst U2 (corresponding to a magnitude limit of $q_{\mathrm{AB}}>19.63$), and the deepest optical-to-radio fluence ratio limit is $F_\text{opt}/F_\text{radio}<0.034$ for burst L82, both in the $q$-band. Compared to previous searches for simultaneous optical emission from other active repeating FRBs, our fluence ratio limit is comparable to that obtained for FRB~20121102A ($\lesssim0.013$ on 0.1\,ms timescales; \citealt{MAGIC18}), but less constraining than that for FRB~20180916B ($\lesssim0.002$ on 10.4\,ms timescales; \citealt{Kilpatrick24}). Our sensitivity to potential fast optical bursts is currently limited by the imaging speed of MeerLICHT. Improving the integration time to milliseconds would probe the fluence ratio limit to $F_\text{opt}/F_\text{radio}\lesssim10^{-6}$ on second timescales for \FRB. We also calculated a summed and averaged strictly simultaneous optical fluence limit per MeerLICHT filter, by calculating the flux limit from each of the co-added ``on'' images. Each flux limit was then  multiplied by the total exposure time of the co-add to obtain a fluence limit, which was then divided by the number of bursts contained in the image. Our deepest summed and averaged optical fluence limit was $F_\text{opt}<0.10$\,Jy\,ms for the $q$-band ``on'' image. The associated optical-to-radio fluence ratio limit was $F_\text{opt}/F_\text{radio}<0.037$. The summed and averaged optical fluence limit is not as deep as the summed and averaged optical fluence limit of $F_\text{opt}<0.029$\,Jy\,ms for FRB~20190520B obtained by \cite{Niino22}. However, the fluence ratio limit of $\lesssim0.037$ is deeper than the summed and averaged optical-to-radio fluence ratio limit of $\lesssim0.20$ obtained by \cite{Niino22}.

\section{Discussion}\label{sec:discussion}

\subsection{Multi-frequency burst rate comparison}

We observed \FRB\ with MeerKAT over a wide frequency range from 544\,MHz to 2.9\,GHz. The contiguous observations at L-band and S-band on 2024 June 26 and the simultaneous observations at UHF and L-band on 2024 June 28 indicate that \FRB\ was most active between 856\,MHz\,--\,1712\,MHz around that time. Assuming the burst rate of \FRB\ is constant in time during our observations, as supported by the comparison between the observed wait-time distribution and the Poissonian distribution (see Section~\ref{sec:burst_rate}), we can characterise the frequency dependence of the burst rate using the following relation:

\begin{equation}
    \frac{\lambda_1}{\lambda_2}=\left(\frac{\nu_1}{\nu_2}\right)^{-\alpha_\text{s}\gamma}\left(\frac{F_{\nu_1,\text{min}}}{F_{\nu_1,\text{min}}}\right)^{\gamma+1},
    \label{eq:spectrum}
\end{equation}

\noindent where $\lambda$ is the burst rate, $F_\text{min}$ is the fluence completeness limit, and $\alpha_\text{s}$ is the  statistical spectral index \citep{Houben19}. Measuring $\alpha_\text{s}$ can be used to understand the statistical distribution of burst energies at different frequencies. Here for simplicity we adopt the same fluence completeness limit of $\sim1$\,Jy\,ms for the UHF, L-band and S-band observations. The burst rates for these three bands above the fluence limit are $\lambda_\text{UHF}=19.5^{+7.2}_{-5.6}\,\text{hr}^{-1}$, $\lambda_\text{L}=50^{+8.7}_{-7.7}\,\text{hr}^{-1}$ and $\lambda_\text{S}=19^{+10.7}_{-7.6}\,\text{hr}^{-1}$, respectively. The errors correspond to the 95\% confidence limit of the Poissonian distribution. Using these values in Eq.~\ref{eq:spectrum} we obtained two values for $\alpha_\text{s}$: $\alpha_\text{s, L/UHF}=1.2^{+0.7}_{-0.6}$ and $\alpha_\text{s, L/S}=-0.9^{+0.6}_{-0.6}$. 
We also compared the burst rate between UHF and L-band using the simultaneous sub-array observations. Given that most of the \FRB\ bursts show narrow-band emission and only 12 bursts were simultaneously detected in both bands (see Section~\ref{sec:morphology}), we simply assumed the bursts detected in each band are independent. That resulted in a spectral index of $\alpha_\text{s, L/UHF}=1.1^{+0.7}_{-0.7}$, consistent with the above value obtained from the full sample.
Despite the large errors on the statistical spectral indices, they suggest a spectral turnover between 856\,MHz--1712\,MHz in the burst energy distribution of \FRB.

There have been measurements of statistical spectral indices for other active repeaters (e.g. \citealt{Houben19, Chawla20, Sand22}). Multiband detections of FRB~20180916B by the GMRT at 400\,MHz and the Green Bank Telescope (GBT) at 800\,MHz were used to derive a spectral index of $\alpha_\text{s}=-0.6^{+1.8}_{-0.9}$ \citep{Sand22}, similar to the value of $\alpha_\text{s, L/S}$ we derived for \FRB\ between 1.3\,GHz--2.4\,GHz. For another repeater, FRB~20121102A, simultaneous observations with the Effelsberg 100\,m radio telescope at 1.4\,GHz and the Low Frequency Array (LOFAR; \citealt{Haarlem13}) at 150\,MHz constrained the spectral index to $>-0.5^{+0.2}_{-0.2}$, suggesting a flattening or turnover of the spectrum at low frequencies. This, combined with the spectral turnover we determined for \FRB, imply there could be similar behaviors in the burst energy spectra of more repeaters. However, it remains unclear whether this varying emission at different frequencies is intrinsic to the emission mechanism or caused by propagation effects such as plasma lensing or free-free absorption in the local environment \citep{Cordes17, Sokolowski18}. Collecting a larger sample of bursts over a broader frequency range from \FRB\ will better constrain the burst energy distribution and possible emission mechanisms.

\subsection{Burst microstructure}

One of the \FRB\ bursts presented in this work consists of many sub-components with durations of $\sim80\,\mu$s, as shown in Figure~\ref{fig:U3}. There could be structures on even shorter timescales, but the limited S/N of the burst prevents us from probing that. In the literature, microstructures have been reported for several repeaters, e.g. FRB~20180916B on timescales of $\sim4\,\mu$s at 1.7\,GHz \citep{Nimmo21}, FRB~20200120E on timescales of $\sim60$\,ns at 1.4\,GHz \citep{Nimmo22} and FRB~20220912A on timescales of $\sim16\,\mu$s at 1.4\,GHz \citep{Hewitt23}. While the microstructure we detected from \FRB\ is not as short as those from other repeaters, the ratio of our observed short timescales to the total burst duration, i.e. $\sim80\,\mu\text{s}/20\,\text{ms}=0.004\ll1$, is comparable to other repeaters. In the context of the FRB being powered by a magnetar, the FRB emission could either originate from the magnetosphere \citep{Kumar17} or the external shock far from the magnetar \citep{Metzger19}. As the external shock model predicts the variability timescale cannot be considerably shorter than the burst duration \citep{Lu22}, the microstructure observed in the \FRB\ burst favors the magnetospheric origin for the FRB emission.

\subsection{Polarisation}

We observe high linear polarisation in most of the MeerKAT-detected repeat bursts from \FRB. The mean linear polarisation fraction is 0.95, comparable to that measured for a few active repeaters such as FRB~20201124A \citep{Jiang22}, FRB~20220912A \citep{Hewitt23} and FRB~20240114A \citep{Tian24b}. 
This value is significantly higher than the mean linear polarisation fraction of 0.63 of the 110-antenna Deep Synoptic Array (DSA-110),  CHIME, and Apertif non-repeating FRB samples \citep{Sherman24, Pandhi24, pastor-marazuela_comprehensive_2025}. 
While there is no clear dichotomy in polarisation properties between repeating and non-repeating FRBs, the active repeaters tend to manifest high linear polarisation more frequently.

\FRB\ also exhibits circular polarisation. Among the MeerKAT-detected bursts, 20\% have $|V/I|>20\%$ with the maximum circular polarisation reaching 52\%. This, combined with previous detections of circular polarisation from all other active repeaters \citep{Hilmarsson21, Feng22b, Jiang22, Kumar22, Xu22, Zhang23b, Xie24}, favors a magnetospheric origin for the FRB emission as the synchrotron maser model involving relativistic shocks does not predict circular polarisation \citep{Metzger19, Plotnikov19, Qu23}. This is further supported by the diversity of PPA variations observed across the pulse profiles of \FRB\ (see Figure~\ref{fig:pol}) that resembles those seen from radio pulsars and magnetars.

\section{Conclusions}\label{sec:conclusions}

In this paper, we report the discovery of the repeating \FRB\ with MeerKAT, and the detection of 249 bursts within 4\,hr of MeerKAT follow-up observations, including 46 in the UHF, 177 in the L-band and 26 in the S-band. This indicates \FRB\ was highly active in June of 2024. Using the voltage buffer data triggered by the bright burst in the L-band, we localised the FRB source to arcsecond precision, facilitating further follow-up observations with other telescopes.

We find the repeat bursts of \FRB\ show frequency downward drift with a mean drift rate of $-16.7\,\text{MHz}\,\text{ms}^{-1}$ (for single-peaked bursts), and are band limited with a mean bandwidth of 169.4\,MHz in the UHF, 287.2\,MHz in the L-band and 423.5\,MHz in the S-band, corresponding to a fractional bandwidth of 0.31, 0.34 and 0.48, respectively. These features are similar to that observed in other known repeaters. We also find microstructure down to $\sim80\,\mu$s in the most complex burst U3 in our sample with no evidence of quasi-periodicity. The limited S/N of this burst prevents us from probing shorter-duration structures as have been observed in other active repeaters.

The fluences of the MeerKAT-detected bursts follow a power-law distribution with an index of $\gamma=-1.6\pm0.1$ and $\gamma=-1.7\pm0.1$ for the UHF and L-band bursts, respectively, above the $\sim1$\,Jy\,ms fluence completeness limit. While the bursts detected in the MeerKAT observations approximately follow a Poissonian repetition, the burst rates in the three bands differ significantly, with the rate in the L-band being $3\times$ higher than in the UHF and S bands. Assuming a constant burst rate for \FRB\ during our observations, we constrain the statistical spectral index among the three bands to $\alpha_\text{s, L/UHF}=1.2^{+0.7}_{-0.6}$ and $\alpha_\text{s, L/S}=-0.9^{+0.6}_{-0.6}$. This suggests a spectral turnover between 856--1712\,MHz in the burst energy distribution of \FRB, which could either be intrinsic to the emission mechanism or induced by propagation effects such as plasma lensing or free-free absorption in the FRB local environment.

We also investigate the polarisation properties of \FRB\ using the triggered voltage buffer data. Most of the bursts are $\sim100\%$ linearly polarised and $\sim10\%\text{--}20\%$ circularly polarised. The maximum circular polarisation reaches 52\%. This, combined with the diversity of PPA variations observed across the \FRB\ bursts, suggests a magnetospheric origin of the FRB emission.

We have searched for optical counterparts of \FRB\ with MeerLICHT contemporaneous optical observations and find no emission, resulting in a fluence upper limit of $F_\text{opt}<0.76$\,Jy\,ms and optical-to-radio fluence ratio limit of $F_\text{opt}/F_\text{radio}<0.034$ on 15\,s timescales in the $q$-band. This demonstrates the capability of MeerLICHT of achieving strictly simultaneous optical observations of a large number of bursts when it comes to following up hyperactive repeating FRBs.

Multiwavelength observations of FRBs can reveal essential properties of the source origins and environments. As a highly active source, \FRB\ is currently being monitored at 700\,MHz--4\,GHz with approximately weekly cadence, and the results will be published in an upcoming paper.

\section*{Acknowledgements}

The authors would like to thank the Director and the operators of MeerKAT and the South African Radio Astronomy Observatory (SARAO) for the prompt scheduling of the observation. 
This project has received funding from the European Research Council (ERC) under the European Union’s Horizon 2020 research and innovation programme (grant agreement No. 694745). JT and BWS acknowledge funding from an STFC Consolidated grant.
IPM acknowledges funding from an NWO Rubicon Fellowship, project number 019.221EN.019. MC acknowledges support of an Australian Research Council Discovery Early Career Research Award (project number DE220100819) funded by the Australian Government. Parts of this research were conducted by the Australian Research Council Centre of Excellence for Gravitational Wave Discovery (OzGrav), project number CE170100004. 
The authors acknowledge useful discussions with Andr\'es G\'urpide.
The MeerKAT telescope is operated by the South African Radio Astronomy Observatory (SARAO), which is a facility of the National Research Foundation, an agency of the Department of Science and Innovation. SARAO acknowledges the ongoing advice and calibration of GPS systems by the National Metrology Institute of South Africa (NMISA) and the time space reference systems department of the
Paris Observatory.The FBFUSE beamforming cluster was funded, installed, and operated by the Max-Planck-Institut fur Radioastronomie and the Max-Planck-Gesellschaft. 


\section*{Data availability}
The MeerKAT data underlying this paper will be shared on reasonable request to the corresponding author.



\bibliographystyle{mnras}
\bibliography{bib, bib_IPM} 



\appendix



\bsp	
\label{lastpage}

\end{document}